\documentclass[12pt]{article}
\usepackage{amsfonts,amssymb,amstext,latexsym,theorem}

\catcode`\@=11
\def\marginnote#1{}
\newcount\hour
\newcount\minute
\newtoks\amorpm
\hour=\time\divide\hour by60
\minute=\time{\multiply\hour by60 \global\advance\minute by-\hour}
\edef\standardtime{{\ifnum\hour<12 \global\amorpm={am}%
        \else\global\amorpm={pm}\advance\hour by-12 \fi
        \ifnum\hour=0 \hour=12 \fi
        \number\hour:\ifnum\minute<10 0\fi\number\minute\the\amorpm}}
\edef\militarytime{\number\hour:\ifnum\minute<10 0\fi\number\minute}
\def\draftlabel#1{{\@bsphack\if@filesw {\let\thepage\relax
   \xdef\@gtempa{\write\@auxout{\string
      \newlabel{#1}{{\@currentlabel}{\thepage}}}}}\@gtempa
   \if@nobreak \ifvmode\nobreak\fi\fi\fi\@esphack}
        \gdef\@eqnlabel{#1}}
\def\@eqnlabel{}
\def\@vacuum{}
\def\draftmarginnote#1{\marginpar{\raggedright\scriptsize\tt#1}}
\def\draft{\oddsidemargin -.5truein
        \def\@oddfoot{\sl preliminary draft \hfil
        \rm\thepage\hfil\sl\today\quad\militarytime}
        \let\@evenfoot\@oddfoot \overfullrule 3pt
        \let\label=\draftlabel
        \let\marginnote=\draftmarginnote
   \def\@eqnnum{(\theequation)\rlap{\kern\marginparsep\tt\@eqnlabel}%
\global\let\@eqnlabel\@vacuum}  }

\def\preprint{\twocolumn\sloppy\flushbottom\parindent 1em
        \leftmargini 2em\leftmarginv .5em\leftmarginvi .5em
        \oddsidemargin -.5in    \evensidemargin -.5in
        \columnsep 15mm \footheight 0pt
        \textwidth 250mmin      \topmargin  -.4in
        \headheight 12pt \topskip .4in
        \textheight 175mm
        \footskip 0pt
        \def\@oddhead{\thepage\hfil\addtocounter{page}{1}\thepage}
        \let\@evenhead\@oddhead \def\@oddfoot{} \def\@evenfoot{} }

\def\titlepage{\@restonecolfalse\if@twocolumn\@restonecoltrue\onecolumn
     \else \newpage \fi \thispagestyle{empty}\c@page\z@ 
        \def\thefootnote{\fnsymbol{footnote}} }

\def\endtitlepage{\if@restonecol\twocolumn \else  \fi
        \def\thefootnote{\arabic{footnote}}
        \setcounter{footnote}{0}}  

\catcode`@=12
\relax

\def\bea{\begin{array}}
\def\bem{\begin{displaymath}}
\def\beq{\begin{equation}}

\def\eea{\end{array}}
\def\eem{\end{displaymath}}
\def\eeq{\end{equation}}

\def\Im{\mathop{\rm Im}}

\def\ov{\overline}

\def\Re{\mathop{\rm Re}}

\def\s2w{\sin^2 \theta_W}
\def\Tr{\mathop{\rm Tr}}

\relax

\def\crbig{\\\noalign{\vspace {3mm}}}

\def\N4{{\cal N}_4}
\def\ds{\displaystyle}

\def\Id{{\rm 1\kern-.28em I}}
\def\Z{{\rm Z\kern-.34em Z}}
\def\R{{\rm I\kern-.2em R}}
\def\C{\hspace{.2em}{\kern-.1em{\raise.47ex\hbox{$\scriptscriptstyle |$}}
\kern-.40em{\rm C}}}
\relax

\renewcommand{\theequation}{\thesection.\arabic{equation}}
\begin{document}
\topmargin-.6cm
%
%
%
%
\begin{titlepage}
\begin{flushright}
NEIP--05--02 \\
hep--th/0502225\\
February/October 2005
\end{flushright}
\vspace{1.1cm}

\begin{center}{\Large\bf
Five-Brane Thresholds and Membrane Instantons 
\\ \vspace{.5cm}
in Four-Dimensional Heterotic M-Theory  }
\vspace{1.5cm}

{\large\bf Luca Carlevaro$^{1}$ and Jean-Pierre Derendinger$^{2}$}
\vskip .5cm
Physics Institute,  Neuch\^atel University, \\
A.-L. Breguet 1, CH--2000 Neuch\^atel, Switzerland
\end{center}
\vspace{1.3cm}

\begin{center}
{\large\bf Abstract}
\end{center}
\begin{quote}
The effective four-dimensional supergravity of M-theory compactified
on the orbifold $S^1/Z_2$ and a Calabi-Yau threefold includes in general 
moduli supermultiplets describing massless modes of five-branes. 
For each brane, one of these fields corresponds to fluctuations along 
the interval. The five-brane also leads to modifications of the 
anomaly-cancelling terms in the eleven-dimensional theory, 
including gauge contributions located on their world-volumes. 
We obtain the interactions of the brane ``interval modulus"  predicted by 
these five-brane-induced anomaly-cancelling terms and we construct their 
effective supergravity description.
In the condensed phase, these interaction terms generate an 
effective non-perturbative superpotential which can also be interpreted 
as instanton effects of open membranes stretching between five-branes
and the $S^1/Z_2$ fixed hyperplanes. Aspects of the vacuum structure of 
the effective supergravity are also briefly discussed.
\end{quote}

\vfill 
\noindent\rule{8.1cm}{0.2mm}\\ 
{\small  
$^1$ {\sl luca.carlevaro@unine.ch} \\
$^2$ {\sl jean-pierre.derendinger@unine.ch }}

\end{titlepage}
\setcounter{footnote}{0}
\setcounter{page}{0}
\newpage
\setlength{\baselineskip}{.6cm}
\setlength{\parskip}{.2cm}

%
%

\section{Introduction}
\setcounter{equation}{0}

Heterotic $E_8\times E_8$ strings compactified to four dimensions on a six-dimensional 
space $K_6$ are also described by M-theory compactified on $K_7 \equiv
S^1/Z_2 \times K_6$ \cite{W, HW}. In particular, it is straightforward to verify 
\cite{NOY, AQ, LOW1, effsugra}
that the effective $\N4=1$ supergravity found in Calabi-Yau or orbifold compactifications 
of perturbative heterotic strings \cite{hetW, DIN2} is reproduced by brane-free M-theory 
configurations with compact space $K_7$. A novelty of the M-theory approach lies
in the possibility to concretely analyse physical effects of non-perturbative brane 
configurations. In the low-energy effective supergravity approximation, configurations 
with five-branes and/or membranes (two-branes) \cite{DLM, M2M5, W} of 
compactified M-theory can be studied from simple modifications of the field 
equations predicted by eleven-dimensional supergravity \cite{CJS}. 

An obvious distinction in the nature of five-brane and membrane effects follows
from the alignement conditions applying to their respective world-volumes
if one requires that the configuration admits (exact or spontaneously broken)  
${\cal N}_4=1$ supersymmetry (four 
supercharges). Each five-brane world-volume is the product of four-dimensional space-time 
and a holomorphic two-cycle in the Calabi-Yau threefold and conditions apply on the respective
cycles of pairs of world-volumes \cite{M2M5, W}. Five-brane massless excitations
\cite{M5modes}, 
which belong to six-dimensional chiral supersymmetry multiplets expanded in modes of 
the two-cycle, lead then to new four-dimensional fields to be included in the effective 
supergravity description. Some of these modes do not depend on the detail of the 
Calabi-Yau geometry: the five-brane modulus describing fluctuations along the $S^1/Z_2$ 
direction, the two-index antisymmetric tensor $\hat B_{\mu\nu}$ with self-dual field
strength and their fermionic ${\cal N}_4=1$ partner. 
These states can be assembled either in a chiral supermultiplet which we will call
$\hat S$ or, in a dual version, in a linear multiplet. The effective supergravity 
for this ``universal five-brane modulus'' supermultiplet has been 
studied in ref.~\cite{DS5brane} (see also ref.~\cite{MPS})\footnote{And, as a function of a 
non-trivial background value of the five-brane modulus, ref.~\cite{LOW, LOW2}.}. 
Firstly, the K\"ahler potential of the 
theory with this new superfield has been obtained and the absence of direct contributions 
to the (perturbative) superpotential has been demonstrated. Secondly, on the basis of the 
four-dimensional superspace structure only, the possible appearance 
of new threshold corrections has been emphasized. 

In contrast, open membrane euclidean world-volumes include the $S^1/Z_2$ direction 
and a cycle in $K_6$ \cite{M2M5, 2bralign}. 
They stretch between the $S^1/Z_2$ fixed planes, or between a fixed plane and a five-brane, 
or between pairs of five-branes. Their effects in the four-dimensional effective supergravity
are then localized in space-time, they can be viewed as instanton-like corrections to the 
interaction Lagrangian. While open membrane stretching between the fixed hyperplanes correspond in the string approach to world-sheet instantons, membranes ending 
on a five-brane describe forces acting on this brane. Their contributions
to the effective supergravity are then expected to lead to new (non-derivative since 
the world-volume includes $S^1/Z_2$) interactions involving the five-brane modulus. 

The corrections to the effective four-dimensional supergravity
induced by the various types of membranes have been studied in 
refs.~\cite{MPS, 2brinst, CK}. They were found to 
contribute to the chiral $F$--density part of the Lagrangian density, in the form of a 
non-perturbative superpotential. Specifically, an interaction bilinear in the
five-brane fermion in superfield $\hat S$ has been computed in 
the four-dimensional background with the five-brane and open membranes ending 
on it. The resulting non-perturbative superpotential shows an exponential dependence 
on the five-brane universal modulus typical of instanton calculus.
To isolate the membrane contributions from other possible non-perturbative sources,
a specific regime is chosen.\footnote{For instance, Moore, Peradze and Saulina 
\cite{MPS} select a
regime where ``open membrane instanton effects are the leading source of non-perturbative effects".} As a consequence, even if the instanton calculation clearly establishes 
the existence of an exponential dependence on $\hat S$, it does not allow to infer how
this exponential term would combine with other non-perturbative contributions which,
like gauge instantons, are expected as 
well.\footnote{Writing the complete non-perturbative 
superpotential as a sum of contributions, as for instance in ref.~\cite{MPS}, is an
assumption which needs to be justified. }

The relevance to physics of the M-theory system with five-branes and membranes
strongly relies upon the structure of superpotentials generated by fluxes, gaugino
condensates and open membrane instantons. Phenomenological questions 
addressed in the literature include supersymmetry breaking and gaugino 
condensation \cite{LOW2, CK}, five-brane stabilization (stabilization
of the modulus $\hat S$) \cite{MPS, CK}, stabilization of all moduli \cite{vacuum},
the existence of stable de Sitter vacua \cite{vacuum, Infl}, inflationaly phases and
potentials \cite{Infl} and cosmic strings \cite{cosmic}. These analyses use
in general simplifying assumptions, in the K\"ahler metric which shows a severe 
mixing of all moduli when five-brane fields are present, or in the superpotential
which is assumed to be a simple sum of non-perturbative contributions. 

In the present paper, we use the anomaly-cancelling terms of the 
eleven-dimensional theory on the orbifold $S^1/Z_2$,
as modified when five-branes are present, to derive new interactions involving the five-brane universal modulus supermultiplet which describes fluctuations along the $S^1/Z_2$ orbifold 
direction. These new interactions are then shown to induce, in the condensed phase, 
the effective non-perturbative instanton superpotential expected from membranes 
stretching between a fixed hyperplane and a five-brane. This superpotential 
correctly reduces to the results of refs.~\cite{MPS, 2brinst, CK} in the regime 
considered in these articles, but its derivation does not require choosing a particular limiting 
regime. This approach provides 
then direct information on the non-perturbative superpotential with combined
gauge and membrane instantons effects.  

The fact that these four-dimensional interactions can be obtained by considering
seemingly unrelated arguments (membrane instanton calculus or gauge 
anomaly-cancelling terms) is a consequence of the superfield structure of the
four-dimensional theory. We use an effective superfield formulation \cite{DSsugra, DS5brane} 
which fully respects the symmetry and supersymmetry ingredients defining 
the microscopic system: the modifications of the Bianchi identities and of
the topological term induced by fixed planes and five-branes. It also respects the
symmetries of the five-brane multiplet, with its self-dual three-form field. 
Thus, a given superspace contribution in the effective Lagrangian describes various
aspects of the microscopic theory, related by the superfield structure of the effective 
theory. This method has been applied to the derivation of the
K\"ahler potential \cite{DS5brane}, including non-linear couplings of the five-brane 
superfield $\hat S$, and we will see in paragraph \ref{secgsquare} that these kinetic 
terms can be derived from (at least) two quite distinct sectors of the microscopic
theory. They can certainly be derived from the Calabi-Yau reduction 
of the five-brane Born-Infeld Lagrangian \cite{DS5brane}. But they can also be derived
from a universal correction to gauge kinetic terms, quadratic in $\hat S$, induced by
$S^1/Z_2$ anomaly-cancellation.  

Schematically, our argument goes as follows. Since we confine ourselves to 
the effective four-dimensional supergravity with up to two derivatives, for which a
(superconformal) superspace formulation exists, counterterms cancelling Lorentz anomalies 
will be irrelevant to our discussion.\footnote{They would however lead to similar phenomena.}
Gauge anomaly-cancelling terms are then entirely due to the ``topological term" of
eleven-dimensional supergravity \cite{CJS}
$$
- \frac{1}{24\kappa^2}\int C_3 \wedge G_4 \wedge G_4.
$$
Sources for the Bianchi identity verified by $G_4$ are provided by the two fixed 
hyperplanes of the $S^1/Z_2$ orbifold and by the aligned five-branes, so that
$$
G_4 = dC_3 + \Delta G_{4, planes} +  \Delta G_{4, branes}.
$$
The contribution $\Delta G_{4, planes}$ depends on the gauge curvatures living on
the planes, and both corrections explicitly depend on the $S^1/Z_2$ coordinate and 
respect the $Z_2$ symmetry used to define the orbifold projection. 
The topological term leads then to a gauge interaction of the form
$$
-{1\over12\kappa^2}\int C_3 \wedge \Delta G_{4, planes} \wedge \Delta G_{4, branes}.
$$
This term gives rise in particular to a gauge interaction of the massless orbifold modes of 
$C_3$ located on the five-brane world-volumes and depending explicitly on their position
along $S^1$. And, after integration over the Calabi-Yau space, it produces a coupling 
to $F\wedge F$ of the axionic partner $\Im T$ of the Calabi-Yau volume 
modulus\footnote{For the bulk moduli $T$ and $S$, we use the terminology familiar
from string compactifications in which $T$ is the volume modulus and $S$ the
dilaton or string coupling modulus. The terminology commonly used in the context of 
$M$-theory, as for instance in refs.~\cite{MPS, LOW}, is unfortunately different. 
Our conventions  are precisely stated in the appendix.} 
$\Re T$ which depends on the five-brane locations along $S^1$. The superfield 
structure developed
for the effective supergravity of the universal five-brane modulus \cite{DS5brane} can then 
be used to understand this interaction as a threshold correction with a calculable 
dependence on the five-brane modulus. In particular, the universal
part of these contributions can be derived from the Dirac-Born-Infeld kinetic Lagrangian.  
The non-perturbative effective superpotential
follows then from standard gaugino condensation. Its dependence on the five-brane 
modulus is precisely the one expected from membrane instanton calculations, 
as performed for instance in ref.~\cite{MPS}.

Along similar lines, a description of some new charged matter contributions arising
from M-theory anomaly-cancellation can
be given. An interesting feature is that the structure organizing 
five-brane threshold corrections is carried over to these matter interactions.

The paper is organised as follows. The anomaly counterterm at the
origin of the five-brane-related gauge interactions is derived in Section \ref{secanom}.
The reduction to four dimensions and the identification of the obtained terms
as superfield densities are then discussed in Sections \ref{secred} and \ref{secsusy}. 
Then, in Section \ref{seccond}, condensation
is performed to derive the non-perturbative superpotential, compare it with 
membrane instanton results and discuss some physical consequences. 
Conclusions and comments are added in Section \ref{secconcl} and an appendix 
collects conventions, notations and technical details. 

\section{The anomaly counterterm}\label{secanom}
\setcounter{equation}{0}

Ten-dimensional hyperplanes located at the $Z_2$ fixed points along $S^1$ and 
five-branes act as sources of the Bianchi identity verified by the four-form field
$G_4$ of eleven-dimensional supergravity. As a consequence of these contributions,
the topological term acquires anomalous variations under local symmetries. 
Together with the variations of the Green-Schwarz term \cite{GS, W, HW, DLM}, 
of order four in the Riemann curvature, these anomalous 
variations are precisely those required to cancel perturbative gauge and 
Lorentz anomalies generated by the $Z_2$ orbifold projection of the 
eleven-dimensional theory and by the chiral gauge multiplets living on the 
hyperplanes. 

The modification of the Bianchi identity and of its solution $G_4$ of course 
leads to modifications of the effective action. All modifications generated by
the topological term would have more than two derivatives in the heterotic, 
ten-dimensional, small $S^1$ radius limit. But compactifying further to four 
dimensions on the Calabi-Yau space $K_6$ also generates modifications
of the effective action at the level of two-derivative gauge terms, because of the 
non-trivial background values of $\langle{\rm tr}\, R^2\rangle$ and 
$\langle{\rm tr}\, F^2\rangle$. The purpose of this section
is to precisely derive some of these terms which arise whenever five-branes
are present. We then begin by recalling some aspects of the description
of $M$-theory on the orbifold $S^1/Z_2$. It should be noted that some ambiguities
remain in our understanding of this description. The gauge
sector relevant to our problem escapes however these ambiguities. 

The explicit formulation of the modified Bianchi identity uses two types of sources, 
associated with hyperplanes supporting Yang-Mills ten-dimensional supermultiplets
and five branes world-volumes. 
On both hyperplanes (labelled by $i=1,2$) live gauge supermultiplets
with curvature two-forms $F_i$. The quantity appearing in the Bianchi identity
is the gauge invariant four-form
\beq
\label{I4is}
I_{4,i} = {1\over(4\pi)^2}\biggl[ {1\over30}{\rm Tr}\, F_i^2 - {1\over2}{\rm tr}\, R^2 \biggr], 
\qquad\qquad 
dI_{4,i}=0, \qquad\qquad (i=1,2),
\eeq
where $R$ is the Lorentz curvature two-form. Five-branes compatible with 
${\cal N}_4=1$ (four supercharges) supersymmetry wrap space-time $M_4$ and
a two-cycle in $K_6$. The transverse Dirac distribution $\delta^{(5)}_{{\cal W}_{6,I}}$
for five-brane number $I$ with world-volume ${\cal W}_{6,I}$ is the five-brane source
in the Bianchi identity, which then reads\footnote{Supersymmetry forbids that
both five-branes and anti-five-branes couple to the $S^1/Z_2$ orbifold.}
\beq
\label{bianchi1}
dG_4 = 2 (4\pi\kappa^2)^{1/3} \biggl[  
{1\over2}  \sum_I  \delta^{(5)}({\cal W}_{6,I}) 
- \sum_i \delta_i I_{4,i} \biggr] .
\eeq
The one-forms 
\beq
\label{deltais}
\delta_1 = \delta(y)\, dy, \qquad
\delta_2 = \delta(y-\pi)\, dy
\eeq
localize the gauge sources on the $Z_2$--fixed hyperplanes.\footnote{The 
appendix collects our conventions.} In order to respect the $Z_2$ symmetry used 
in the orbifold
projection, we actually label with index $I$ {\it pairs} of five-branes located at 
$\pm y_I$.\footnote{And one may then choose $0 \leq y_I \leq \pi$.}

The procedure to resolve ambiguities in the solution of the Bianchi identity
has been discussed in detail in the literature.\footnote{For instance in refs.~\cite{BDS}
and \cite{BM}, and references therein.} The general solution of eq. (\ref{bianchi1}) includes several arbitrary integration constants which are constrained by consistency conditions: the four-form
field $G_4$ should be gauge-invariant and globally well-defined, its action should be 
well-defined, the fields eliminated by the $Z_2$ truncation should be gauge-invariant,
as well as the massive modes of the $S^1$ expansion of the surviving states. 
These conditions should be verified for any number of five-branes at 
arbitrary locations. Writing
\beq
\label{G41}
G_4 = d C_3 + \Delta G_{4, branes} + \Delta G_{4, planes},
\eeq
the consistency constraints point to a unique solution for the hyperplane 
contribution\footnote{This is the solution with ``$b=1$" in refs.~\cite{BDS, BM}.}:
\beq
\label{G4planes}
\begin{array}{rcl}
\Delta G_{4,planes} &=& -(4\pi\kappa^2)^{1/3}\sum_i \bigl(
\epsilon_i \, I_{4,i} - {dy\over\pi} \wedge \omega_{3,i} \bigr),
\crbig
&=& -(4\pi\kappa^2)^{1/3}\sum_i \bigl(
d[\epsilon_i \, \omega_{3,i}] - 2\delta_i \wedge \omega_{3,i} \bigr),
\end{array}
\eeq
where, as discussed in detail in the appendix,
\beq
\label{epsilonis1}
\epsilon_i(y) = {\rm sgn}\, (y-y_i) - {y-y_i \over \pi}, \qquad
y_1=0, \qquad y_2=\pi, \qquad y \in [-\pi, \pi]
\eeq
and the Chern-Simons three-forms are defined by 
\beq
\label{omegais}
d\omega_{3,i} = I_{4,i}, \qquad\qquad i=1,2. 
\eeq

A similar discussion can be made for the five-brane contribution to $G_4$.
As already mentioned, the five-branes are space-time filling and wrap a 
holomorphic two-cycle in $K_6$. This implies that one can certainly write
\beq
\label{delta5}
\delta^{(5)} ({\cal W}_{6,I}) = [\delta(y-y_I) + \delta(y+y_I)] 
dy \wedge \delta^{(4)} ({\cal W}_{6,I}), 
\eeq
where $\delta^{(4)} ({\cal W}_{6,I})$ is now a four-form Dirac distribution in 
$K_6$ such that 
$$
\int_{M_4\times K_6} I_6 \wedge  \delta^{(4)} ({\cal W}_{6,I})
= \int_{{\cal W}_{6,I}} I_6
$$
for any six-form $I_6$. The natural solution of the Bianchi identity is then 
to include in $G_4$ the brane contribution
\beq
\label{G4branes}
\begin{array}{rcl}
\Delta G_{4,branes} &=& (4\pi\kappa^2)^{1/3} \sum_I  \left[{1\over2}
\hat\epsilon_{y_I}(y) \delta^{(4)}({\cal W}_{6,I}) 
- {1\over\pi}dy \wedge \theta^{(3)}({\cal W}_{6,I}) \right] + d{\cal C}_3
\crbig
&=&  - (4\pi\kappa^2)^{1/3} \sum_I  
\bigl\{ \delta(y-y_I) + \delta(y+y_I) \bigr\} dy \wedge \theta^{(3)}({\cal W}_{6,I})
\crbig
&& + d \bigl\{ {\cal C}_3 + {1\over2} (4\pi\kappa^2)^{1/3} \sum_I 
\hat\epsilon_{y_I}(y) \theta^{(3)}({\cal W}_{6,I}) \bigr\} ,
\end{array}
\eeq
where $d \theta^{(3)}({\cal W}_{6,I}) =  \delta^{(4)}({\cal W}_{6,I})$ and
\beq
\label{epsilon2}
\begin{array}{l}
\hat\epsilon_{y_I}(y) = {\rm sgn}\, (y - y_I) + {\rm sgn}\, (y + y_I) - {2y\over \pi},
\qquad\qquad 0\leq y_I \leq \pi,
\crbig
d\hat\epsilon_{y_I}(y) = 2\left[\delta(y-y_I) + \delta(y+y_I) - {1\over\pi}\right]dy.
\end{array}
\eeq
Notice that $\epsilon_1(y) = {1\over2}\hat\epsilon_{0}(y)$ and
$\epsilon_2(y) = {1\over2}\hat\epsilon_{\pi}(y)$. The addition in eq. (\ref{G4branes}) 
of the term $d{\cal C}_3$ allows for the introduction of brane modes contributions
into the topological term, if necessary \cite{BM}: it is known \cite{5br, W5br} that a 
gauging by $C_3$ of the 
three-form field of five-brane fluctuations is induced by consistent coupling of 
five-branes to eleven-dimensional supergravity. We will not elaborate on this point 
here since in four dimensions the needed terms would automatically appear in the 
supersymmetrization of the effective theory.\footnote{See 
Section \ref{secsusy}.} 

To summarize, one can write
\beq
\label{G42}
\begin{array}{rcl}
G_4 &=& d\tilde C_3 + 2(4\pi\kappa^2)^{1/3}\sum_i \delta_i \wedge \omega_{3,i} 
\crbig
&& - (4\pi\kappa^2)^{1/3} \sum_I  
\bigl\{ \delta(y-y_I) + \delta(y+y_I) \bigr\} dy \wedge \theta^{(3)}({\cal W}_{6,I}),
\crbig
\tilde C_3 &=& C_3 -(4\pi\kappa^2)^{1/3}\left[\sum_i  \epsilon_i \, \omega_{3,i} 
- {1\over2} \sum_I  \hat\epsilon_{y_I}(y) \theta^{(3)}({\cal W}_{6,I}) \right]
+ {\cal C}_3 ,
\end{array}
\eeq
and $G_4$ and $d\tilde C_3$ only differ at locations where 
hyperplanes or five-branes sit. 
 
The gauge transformation of $C_3$ is completely fixed by consistency
of the $Z_2$ orbifold projection. The topological term indicates that $C_3$ is
intrinsically odd. The components $C_{ABC}$, $A,B,C=0.\ldots,9$, are then
projected out and should then for consistency be gauge invariant. This 
condition implies
\beq
\label{deltaC3}
\begin{array}{rcl}
\delta C_3 &=& (4\pi\kappa^2)^{1/3}{1\over\pi} \, dy \wedge \sum_i \omega_{2,i}^1 \,,
\crbig
\delta\tilde C_3 &=& (4\pi\kappa^2)^{1/3} \sum_i\left[ 2\delta_i\wedge  \omega_{2,i}^1
-d\{ \epsilon_i \omega_{2,i}^1 \} \right],
\end{array}
\eeq
where $\delta\omega_{3,i} = d\omega_{2,i}^1$. The correct modified topological 
term, which cancels anomalies locally, is then \cite{BM}
\beq
\label{topol}
- {1\over24\kappa^2} \int \tilde C_3\wedge G_4\wedge G_4.
\eeq

Our goal is to infer from this modified topological term four-dimensional
interactions of the massless $S^1/Z_2 \times K_6$ modes. The substitution
of eqs. (\ref{G42}) leads to two classes of terms. Firstly, contributions
involving the massless mode $C_{ABy}$ ($A,B=0,\ldots,9$).
This mode corresponds to the antisymmetric tensor $B_{AB}$ of 
ten-dimensional sixteen-supercharge supergravity. From gauge-transformation
(\ref{deltaC3}), one deduces that the appropriate definition with 
$\delta B = (4\pi)^2 \sum_i \omega_{2,i}^1$ is\footnote{Taking into account the factor 
$(4\pi)^{-2}$ in the definitions of $I_{4,i}$ and $\omega_{3,i}$. This is the definition
of the two-form field commonly used in ten dimensions, with dimension 
(mass)$^2$.} 
\beq
\label{Bis}
B_{AB} = (4\pi\kappa^2)^{-1/3} (4\pi)^2 \, \pi R_{S^1} \, C_{ABy}^{(0)}, 
\qquad
C_{ABy}^{(0)} = {1 \over 2\pi} \int_{-\pi}^\pi dy\, C_{ABy},
\eeq
where $R_{S^1}$ is the $S^1$ radius.
In terms of this massless field, the topological term produces the 
following four-dimensional interactions
\beq
\label{topol3}
\displaystyle{ -{1\over32\pi^2} \int_{S^1 \times K_6}
B_2 \wedge\biggl[ \sum_{i,j}   \epsilon_i \, \epsilon_j \, I_{4,i}\wedge I_{4,j} 
- \sum_{i,I}  \epsilon_i \, \hat\epsilon_{y_I} \, I_{4,i} \wedge
\delta^{(4)}({\cal W}_{6,I}) \biggr], }
\eeq
with $\int_{S^1} = \int_{-\pi}^\pi dy$.
The first contribution, when integrated over $S^1$ only, generates the 
Green-Schwarz gauge anomaly-cancelling terms expected for the $E_8\times
E_8$ heterotic string \cite{GS}. Its consequences for Calabi-Yau compactifications 
have been studied long ago \cite{IN, DIN2}.
When integrated over $S^1\times K_6$, it leads to 
gauge threshold corrections depending on the Calabi-Yau volume modulus
as well as dilaton-dependent charged-matter terms. The second contribution, 
which also depends on the $S^1$ position of the five-branes, is of interest to us. 

The modified topological term also produces the following $C_3$--independent 
gauge interactions:
\beq
\label{topol2gauge}
\begin{array}{l}
\displaystyle
-{1\over3} \int_{S^1\times K_6} \biggl[ 
-\sum_{i,j,k}\epsilon_i \epsilon_j \, \omega_{3,i} \wedge I_{4,j} \wedge
\omega_{3,k }
+{1\over2}\sum_{i,J,k} \epsilon_i \hat \epsilon_{y_J} \,
\omega_{3,i} \wedge \delta^{(4)} ({\cal W}_{6,J}) \wedge  \omega_{3,k} 
\crbig \hspace{1.9cm} \displaystyle
+{1\over2}\sum_{i,J,k}  \epsilon_i \hat \epsilon_{y_J} 
I_{4,i} \wedge \theta^{(3)} ({\cal W}_{6,J}) \wedge  \omega_{3,k} 
+ \sum_{i,j,K}  \epsilon_i  \epsilon_j 
\omega_{3,i} \wedge I_{4,j} \wedge \theta^{(3)} ({\cal W}_{6,K}) \biggr] .
\end{array}
\eeq
After integration over $y$, the first contribution is a local counterterm
allowed by anomaly cancellation \cite{BM}\footnote{The anomaly twelve-form
obtained from descent equations vanishes.}.
The next three terms are non-trivial
brane contributions depending on their positions along $S^1$. 

The overlap integrals over $S^1$ give an interesting result. First introduce the
numbers
\beq
\label{int1}
d_{ij}\,=\,\int_{-\pi }^{\pi }dy\,\epsilon _{i}\,\epsilon _{j}\,=\,\frac{\pi 
}{3}\,(3\delta _{ij}-1)
\eeq
for the first integrals in contributions (\ref{topol3}) and (\ref{topol2gauge}).
For those depending on the brane position, define the variables 
\beq
\label{Deltadef}
\Delta_{I,1} = { y_I \over \pi} ,
\qquad
\Delta_{I,2} = 1- { y_I \over \pi} ,
\qquad\qquad \Bigl(
0 < y_I < \pi; \quad 
0 < \Delta_{I,j} < 1 \Bigr), 
\eeq
the distances from brane $I$ to the fixed planes, with normalization
$\Delta_{I,1}+\Delta_{I,2}=1$. Then, 
\beq
\begin{array}{rcccl}
\displaystyle
{1\over2\pi} \int_{-\pi}^\pi dy\, \epsilon_1(y)\hat\epsilon_{y_I}(y)
&=& \displaystyle (\Delta_{I,2})^2 - {1\over3} 
&=& \displaystyle {y_I^2\over\pi^2} - 2{y_I\over\pi} + {2\over3},  
\crbig \displaystyle
{1\over2\pi} \int_{-\pi}^\pi dy\, \epsilon_2(y)\hat\epsilon_{y_I}(y)
&=& \displaystyle (\Delta_{I,1})^2 - {1\over3}
&=& \displaystyle {y_I^2\over\pi^2} - {1\over3}.
\end{array}
\eeq
It will be important for the supersymmetrization of the four-dimensional interactions 
that the terms quadratic in $y_I$ are identical in both integrals. With these results,
contributions (\ref{topol3}) lead to
\beq
\label{topol4}
\begin{array}{l}
\displaystyle{ -{1\over16\pi} \int_{K_6}
B_2 \wedge\biggl[ {1\over3}( I_{4,1}^2 + I_{4,2}^2 - I_{4,1}I_{4,2}) }
\crbig \hspace{1.8cm}
\displaystyle{ - \sum_I  \delta^{(4)}({\cal W}_{6,I}) \wedge
\Bigl\{ \Bigl(\Delta_{I,2}^2 - {1\over3} \Bigr) I_{4,1} 
+ \Bigl(\Delta_{I,1}^2 - {1\over3} \Bigr) I_{4,2} 
\Bigr\}\Bigr]  , }
\end{array}
\eeq
while expressions (\ref{topol2gauge}) give
\beq
\label{topol5}
\begin{array}{l}
\displaystyle{
{\pi\over3} \, \int_{K_6} 
\biggl \{ \omega_{3,1} \wedge  \omega_{3,2} \wedge (I_{4,1}-I_{4,2}) }
+{3\over2} \sum_I ( \Delta_{I,1} - \Delta_{I,2} ) \delta^{(4)} ({\cal W}_{6,I})
\wedge \omega_{3,1} \wedge\omega_{3,2} 
\crbig \hspace{15mm}
\displaystyle{ 
- \sum_I  \theta^{(3)}({\cal W}_{6,I}) \wedge \Bigl[
( \Delta_{I,2} \, \omega_{3,1} - \Delta_{I,1} \, \omega_{3,2} ) \wedge 
( \Delta_{I,2} \, I_{4,1} - \Delta_{I,1} \, I_{4,2} ) }
\crbig \hspace{38mm}
\displaystyle{
- \omega_{3,1} \wedge I_{4,1}  
- \omega_{3,2} \wedge I_{4,2} 
+ {1\over2} \omega_{3,1} \wedge I_{4,2}  
+ {1\over2} \omega_{3,2} \wedge I_{4,1}  \biggr] \biggr\} , }
\end{array}
\eeq
after some partial integrations. 

The anomalous gauge variation of contributions (\ref{topol4}) and (\ref{topol5}) 
can be written as
\beq
\label{gaugevar}
\begin{array}{l} \displaystyle
-{\pi\over3} ( \omega_{2,1}^1\, I_{4,1}^2 +  \omega_{2,2}^1 \, I_{4,2}^2)
\crbig \displaystyle
+ {2\pi\over3} \sum_I  \delta^{(4)} ({\cal W}_{6,I}) \wedge 
[\Delta_{I,2} \, \omega_{2,1}^1 - \Delta_{I,1} \, \omega_{2,2}^1]
\wedge [ \Delta_{I,2} \, I_{4,1}  - \Delta_{I,1} \, I_{4,2}].
\end{array}
\eeq
Applying descent equations to these variations leads to the formal
anomaly twelve-form
\beq
\label{12form}
\begin{array}{rcl}
{\cal I}_{12} &=& 
-{\pi\over3} \Bigl[ I_{4,1} + I_{4,2} -  \sum_I  \delta^{(4)} ({\cal W}_{6,I}) \Bigr]
\crbig &&
\wedge \Bigl[ I_{4,1}^2 + I_{4,2}^2 - I_{4,1}I_{4,2} 
+ \sum_I  \delta^{(4)} ({\cal W}_{6,I}) \wedge
\{ ( 1 - 3\Delta_{I,2}^2 )I_{4,1} + ( 1 - 3\Delta_{I,1}^2)I_{4,2}
\}  \Bigr]
\crbig
&=& {\cal I}_{12, het.} +  \sum_I  \delta^{(4)} ({\cal W}_{6,I}) \, {\cal I}_{8, I} \, ,
\end{array}
\eeq
since for four-dimensional space-time-filling five-branes,
$\delta^{(4)} ({\cal W}_{6,I}) \wedge\delta^{(4)} ({\cal W}_{6,J}) =0$.
The contribution of each five-brane is encoded in the eight-form
\beq
\label{brane8form}
{\cal I}_{8, I} = \pi [ \Delta_{I,2} I_{4,1}  - \Delta_{I,1} I_{4,2}]^2,
\eeq
while the heterotic contribution is as usual 
${\cal I}_{12, het.} = - {\pi\over3} [ I_{4,1}^3 + I_{4,2}^3]$.

The form of ${\cal I}_{8,I}$, eq. (\ref{brane8form}), clearly shows that the distance 
from the brane to the first plane acts as the (inverse squared)
coupling of the gauge fields living on the second plane. Similarly,
the distance from the brane to the second plane acts as the (inverse squared)
coupling of the gauge fields living on the first plane.

At this point, the conclusion is that the effective, four-dimensional supergravity 
includes gauge contributions due to five-branes which arise from
the Calabi-Yau reduction of expressions (\ref{topol4}) and (\ref{topol5}), as
derived from the modified topological term (\ref{topol}). In the next two sections, 
we perform this reduction keeping only the ``universal" massless modes 
unrelated to geometrical details of $K_6$, but including the five-brane modulus
along the $S^1$ direction, and we write the effective 
four-dimensional supergravity using superconformal tensor calculus. 

\section{Reduction to four dimensions}\label{secred}
\setcounter{equation}{0}

In the reduction to four dimensions, we use the freedom to rescale moduli fields 
to set the $S^1$ circumference $2\pi R_{S^1}$ and the Calabi-Yau volume 
$V_6$ to unity.\footnote{The four-dimensional gravitational constant is then 
$\kappa_4^2 = \kappa^2$, but we nevertheless
use different symbols since their mass dimensions differ.}

As usual, the massless modes of the metric tensor expanded on $M_4 \times K_7$ 
include $g_{\mu\nu}$, the scalar field $g_{yy}$ and massless modes of the internal
metric $g_{i\ov k}$. Among these, we only keep the universal, K\"ahler-metric 
volume modulus. Similarly, the antisymmetric tensor $C_{ABy}$ leads
to a massless $B_{\mu\nu}$ and we only keep
the universal massless mode of the internal tensor $B_{i\ov k}$. With these bosonic modes and their fermionic partners, the reduction
of eleven-di\-men\-sio\-nal supergravity can be described by two chiral multiplets
$S$ and $T$ with the familiar K\"ahler potential \cite{hetW, DIN2}\footnote{In general, we use 
the same notation for a chiral supermultiplet and for its lowest complex scalar 
component.}
\beq
\label{usualK}
{\cal K} = - \ln(S + \ov S) - 3\ln ( T + \ov T ).
\eeq
Following eq. (\ref{Bis}), we define 
\beq
\label{termB}
B_{\mu \nu }= {(4\pi )^2 \over 2} (4\pi \kappa ^{2})^{-1/3} \, C_{\mu \nu
y}\,,
\qquad
B_{i\bar{j}}= {i\over\kappa_4^2} \, \Im T \, \delta _{i\bar{j}}
\eeq
and $B_{\mu\nu}$ is dual to $\Im S$.  

With five-branes, vector fields on the two fixed hyperplanes gauge an algebra
$G_1 \times G_2$ which is further broken by the Calabi-Yau compactification. 
Embedding the $SU(3)$ holonomy into $G_1 \times G_2$ defines the 
four-dimensional gauge group $G_{(4)}$ as the stabilizer of this $SU(3)$
in $G_1 \times G_2$. Calabi-Yau reduction of the ten-dimensional gauge
fields\footnote{We find useful to keep track of the plane index $i=1,2$ and 
$\alpha (i)$ is then an index in the adjoint representation of $G_i$. Similarly,
$a(i)$ will be used for the adjoint of $G_{(4)}$ and $m(i)$ for the representation
spanned by complex scalar fields. } 
${\cal A}_B^{\alpha (i)}$ leads then to the corresponding gauge fields 
${\cal A}_\mu^{a (i)}$. It also produces a set of $SU(3)$--singlet complex 
scalar fields ${\cal A}^{m (i)}$ in some representation of $G_{(4)}$. 

With up to two derivatives, Riemann curvature contributions in counterterms
(\ref{topol4}--\ref{topol5}) can be omitted. The Calabi-Yau reduction of $I_{4,i}$ 
delivers then:
\beq
\label{termI}
\begin{array}{rcl}
I_{i, \mu \nu \rho \sigma} &=& \displaystyle{
\frac{3!}{(4\pi )^{2}}\, \sum_\alpha
F^{\alpha(i)}_{[\mu \nu} F^{\alpha(i)}_{\rho \sigma ]} 
\,,}
\crbig
I_{i, \mu \nu k\bar{l}} &=& \displaystyle{
-\frac{4}{(4\pi )^{2}}\sum_m 
(\mathcal{D}_{[\mu }  {\cal A}^{m (i)})
(\mathcal{D}_{\nu ]} \ov {\cal A}^{m (i)})  \,\delta _{k\bar{l}} }
\crbig 
&=& \displaystyle{
 \frac{2i}{(4\pi )^{2}}\sum_m \partial _{[\mu } \left(
 {\cal A}^{m (i)} \mathcal{D}_{\nu ]} \ov {\cal A}^{m (i)}
 -  \ov {\cal A}^{m (i)} \mathcal{D}_{\nu ]} {\cal A}^{m (i)} 
 \right) \,i \delta _{k\bar{l}} } \,,
 \crbig
I_{i, \mu jkl} &=& \displaystyle{
{2 \over(4\pi)^2} \, \partial _{\mu }\left(
\lambda _{i, mnp} {\cal A}^{m (i)}{\cal A}^{n (i)}{\cal A}^{p (i)} \right)
\epsilon _{jkl} \,. }
\end{array}
\eeq
In the last expression,  $\lambda _{i, mnp}$ is the symmetric tensor invariant under
$G_{(4)}$ that may arise from the internal Chern-Simons term $\omega_{i,\, jkl}$.
We will use the notations
\beq
\lambda{\cal A}^3 = \sum_i \lambda{\cal A}_i^3,
\qquad 
\lambda{\cal A}_i^3  \equiv 
 \lambda _{i, mnp} {\cal A}^{m (i)}{\cal A}^{n (i)}{\cal A}^{p (i)} 
\eeq
to denote this cubic holomorphic couplings which also appear in the
superpotential 
\beq
\label{Wis1}
W = c + \lambda{\cal A}^3.
\eeq
Finally, $I_{4,i}$ has a non-trivial background value 
$\langle I_{4,i}\rangle_{ij\ov k\ov l}$.

With these results, the reduction to four dimensions of the first line in expression
(\ref{topol4}), which depends on $B_2$ and exists without five-brane 
can be written
\beq
\label{Lplane}
\begin{array}{rcl}
\Delta \mathcal{L}_{plane}  &=&  \displaystyle{ \frac{1}{2(4\pi )^4}
\,\sum_{i,j}d_{ij} \biggl\{ \frac{1}{\kappa_4 ^{2}} \, e_4 \,    
\langle \mathcal{I}_{j}\rangle \,  \Im T \,
F^{\alpha(i)}_{\mu \nu}\widetilde{F}^{\alpha(i)\,\mu \nu}    }
\crbig
&& \displaystyle{ - i \langle\mathcal{I}_{j}\rangle} \,
\epsilon^{\mu \nu \rho \sigma} (\partial _\mu B_{\nu \rho })
\sum_{m}[{\cal A}^{m(i)}(\mathcal{D}_{\sigma }\ov {\cal A}^{m(i)})\,-\,
\ov  {\cal A}^{m(i)}(\mathcal{D}_{\sigma }{\cal A}^{m(i)})]
\crbig
&& \displaystyle{
- \frac{i}{(4\pi )^2} \epsilon^{\mu \nu\rho \sigma }
(\partial _{\mu }B_{\nu \rho }) \,
(\lambda{\cal A}_i^3\,\partial_{\sigma}(\overline{\lambda{\cal A}}_j^3)
-\partial_{\sigma}(\lambda{\cal A}_i^3)\,\overline{\lambda{\cal A}}_j^3)
\biggr\} .  }
\end{array}
\eeq
The background value of $I_{4,i}$ is encoded in the
integral over the Calabi-Yau manifold
\beq
\label{Iiis}
\langle \mathcal{I}_{i}\rangle = V_6^{-1}
\int_{K_{6}} \,\langle I_{i}\rangle_{kl\overline{kl}} \,
\delta _{j\overline{j}}\,\epsilon^{jkl}\epsilon^{\overline{jkl}}. 
\eeq
In expression (\ref{Lplane}), the first term is a threshold correction depending
on the volume modulus already well-known from the heterotic strings \cite{DIN2, IN}.
The second and third contributions describe interactions of matter scalars 
with the string coupling multiplet. They have been considered in detail in refs.
\cite{NOY, LOW1, DSsugra}. 

The reduction of the five-brane contribution in expression (\ref{topol4})
leads to the following Lagrangian terms:
\beq
\label{Lbrane}
\begin{array}{rcl}
\Delta \mathcal{L}_{brane} &=& \displaystyle{
\frac{1}{8(4\pi)^3} \sum_{I,i}  a_I
\Bigl( \Delta _{I,i}^2 - {1\over3} \Bigr)
\,\Bigl[ \frac{1}{\kappa_4 ^{2}} \, e_4
\,\Im T \, F^{\alpha(i)}_{\mu \nu}\widetilde{F}^{\alpha(i)\,\mu \nu}   }
\crbig
&& \hspace{2.2cm}\displaystyle{ 
-  i \,  \epsilon^{\mu \nu \rho \sigma }\partial_\mu B_{\nu \rho}
\sum_m [ {\cal A}^{m(i)}(\mathcal{D}_{\sigma }\ov{\cal A}^{m(i)})\,-\,
\ov{\cal A}^{m(i)}(\mathcal{D}_{\sigma }{\cal A}^{m(i)})]\Bigr] .  }
\end{array}
\eeq
The area of the Calabi-Yau two-cycle (with coordinate $z$) wrapped by the 
five-brane world-volume can be written
\beq
\label{area}
{ a}_I = \int_{\mathcal{C}_{2,I}}dz\,d\overline{z}\,\frac{\partial z^{m}}{\partial z}%
\frac{\partial \overline{z}^{\overline{n}}}{\partial \overline{z}}\,\delta
_{m\overline{n}} \, .
\eeq
The first term in (\ref{Lbrane}) is the five-brane contribution to gauge
threshold corrections. Its existence has been demonstrated in an 
explicit background calculation by Lukas, Ovrut and Waldram \cite{LOW, LOW2}. 
The second term is again a matter interaction with the string coupling multiplet.
Both terms depend on the positions $y_I$ of the five-branes. Hence, they depend
on the $S^1/Z_2$ five-brane modulus. 

The terms collected in expression (\ref{topol5}) are somewhat ambiguous since
they are defined up to contributions which, like the first one or any counterterm of the form $\theta^{(3)}({\cal W}_{6,I})\wedge I_7$, do not contribute to the 
gauge-invariant anomaly twelve-form. To reduce the first term, introduce the 
four-dimensional Chern-Simons forms
\beq
\begin{array}{l}
{\ds \partial_{[\mu}\omega_{i,\nu \rho \sigma]} = {1\over 4}\,F_{i,[\mu\nu}\, F_{i,\rho \sigma]}\,, \qquad \qquad 
\partial_{[\mu}\omega_{i,\nu]j\bar k} = \,{1\over 2}  \,I_{i,\mu \nu j\bar k}\,,}
\crbig
\partial_{\mu}\omega_{i,jkl} =  \,I_{i,\mu jkl}\,.
\end{array}
\eeq
The first term then generates couplings of charged matter scalars to gauge fields:
\beq
\label{Lplane2}
\begin{array}{rcl}
\Delta {\cal L}_{plane}  &=&  {\ds \frac{i}{3(4\pi)^5}
\epsilon^{\mu \nu \rho \sigma} \omega_{1,\mu \nu \rho}\,\Big[
\lambda{\cal A}_2^3\; \partial_{\sigma}(\overline{\lambda{\cal A}}_2^3)
-\overline{\lambda{\cal A}}_2^3\; \partial_{\sigma}(\lambda{\cal A}_2^3)}
+\,\lambda{\cal A}_1^3\; \partial_{\sigma}(\overline{\lambda{\cal A}}_2^3)
\crbig
&&
-\overline{\lambda{\cal A}}_1^3\; \partial_{\sigma}(\lambda{\cal A}_2^3)
\,+\,2[\overline{\lambda{\cal A}}_2^3\; \partial_{\sigma}(\lambda{\cal A}_1^3)
-\lambda{\cal A}_2^3\; \partial_{\sigma}(\overline{\lambda{\cal A}}_1^3)] \Big] +
(1\leftrightarrow 2)
\crbig
&&
- {\ds \frac{i}{12(4\pi)^3}
\epsilon^{\mu \nu \rho \sigma}\,\omega_{1,\mu \nu \rho}}
{\sum_{m}({\cal A}^{m(2)}\mathcal{D}_{\sigma }\ov {\cal A}^{m(2)}-
\ov  {\cal A}^{m(2)}\mathcal{D}_{\sigma}{\cal A}^{m(2)})\, 
(\langle {\cal I}_1\rangle - \langle {\cal I}_2\rangle)}
\crbig
&& 
+ (1\leftrightarrow 2) .
\end{array}
\eeq 
As we will see in the next section, these terms do not have a natural
supersymmetric extension in general, a fact which may have some relation to
their ambiguous character. Notice however that in the minimal embedding of 
the Calabi-Yau background into one plane only, most of these mixing terms are 
absent and a natural supersymmetrization exists. 

Likewise, the second term in eq. (\ref{topol5}) yields gauge--matter
interactions depending on the five-brane positions along $S^1$:
\beq
\label{Lbrane2}
\begin{array}{rcl}
\Delta \mathcal{L}_{brane} &=& \displaystyle{
 \frac{i}{8(4\pi)^3} \sum_{I}  a_I\, \left( \Delta _{I,1} - \Delta _{I,2} \right)
\,\epsilon^{\mu \nu \rho \sigma}
 \omega_{1,\mu\nu \rho} }
\crbig
&& \hspace{2.2cm}\displaystyle{
\times \sum_m [ {\cal A}^{m(2)}(\mathcal{D}_{\sigma }\ov{\cal A}^{m(2)})\,-\,
\ov{\cal A}^{m(2)}(\mathcal{D}_{\sigma }{\cal A}^{m(2)})] }
+ (1\leftrightarrow 2) .
\end{array}
\eeq

In the next section, we will derive the effective four-dimensional supergravity 
in the particular case of a single five-brane. To simplify, we will omit 
the index $I$ and the corresponding sums.
We will however use a formulation in which restoring contributions of several
branes is straightforward. 

\section{The effective supergravity}\label{secsusy}
\setcounter{equation}{0}

The universal $S^1/Z_2$ five-brane modulus describing fluctuations along the 
interval direction has a supersymmetric bosonic partner arising from the mode 
expansion of the five-brane antisymmetric tensor $\hat B_{mn}$. Six-dimensional 
(world-volume) supersymmetry of the five-brane multiplet requires that the 
three-form curvature $\hat H_{mnp}$ of this tensor is self-dual. For the massless
four-dimensional universal mode, self-duality is the 
condition\footnote{Omitting fermion and covariantization contributions.}
\beq
\label{asd}
\hat H_{\mu\nu\rho} \equiv 3\partial_{[\mu}\hat B_{\nu\rho]} = e_4 
{\epsilon_{\mu\nu\rho}}^\sigma \partial_\sigma\hat B_{i\ov j} 
\equiv e_4 {\epsilon_{\mu\nu\rho}}^\sigma 
\hat H_{\sigma i\ov j}.
\eeq
Then, clearly, the four-dimensional supersymmetric description of the five-brane 
universal modulus uses either a linear multiplet $\hat L$ with the tensor 
$\hat H_{\mu\nu\rho}$ and a real scalar $\hat C$ for the modulus, 
or a chiral multiplet with complex scalar $\hat S$ and $\Im \hat S$ related to
$\hat B_{i\ov j}$. The supersymmetric extension of condition (\ref{asd}) is 
chiral-linear duality, the duality transformation exchanging superfields $\hat S$ 
and $\hat L$ \cite{Slinear}. 

The fact that the chiral multiplet $\hat S$ is dual to a linear multiplet has three 
important consequences for its supergravity couplings. Firstly, the K\"ahler potential 
is a function of $\hat S+\ov{\hat S}$ only. Secondly, the holomorphic
gauge kinetic function can only depend linearly on $\hat S$. These two 
consequences follow from the intrinsic gauge invariance of $\hat B_{\mu\nu}$, 
which translates into axionic shift symmetry of $\hat S$ in the chiral formulation. 
Thirdly, the superpotential does not depend on $\hat S$. In supergravity, in 
contrast to global supersymmetry, this statement is ambiguous and directly linked 
to the first consequence above. The superpotential $W$
and the K\"ahler potential ${\cal K}$ are not independent: the entire theory depends only 
on ${\cal G} = {\cal K} + \ln|W|^2$. Terms can then be moved from or into ${\cal K}$ 
provided they are harmonic functions of the complex chiral fields. 
The correct statement is then that ${\cal G}$ may only depend on 
$\hat S+\ov{\hat S}$. Moving terms from ${\cal K}$ to the superpotential can 
artificially generate a dependence on $\hat S$ of the form 
\beq
\label{Ssuperpot}
W_{new} = e^{b\hat S} W (z^i),
\eeq
where $b$ is a real constant and $z^i$ denotes all other chiral multiplets, and 
a new K\"ahler potential ${\cal K}_{new} (\hat S+\ov{\hat S},z^i, \ov z_i)$
such that the function ${\cal G}$ remains unchanged,
${\cal K} + \ln|W|^2 = {\cal K}_{new} + \ln|W_{new}|^2$. Notice that adding a 
$\hat S$--independent term to the superpotential (\ref{Ssuperpot}) is not allowed. 
Non-perturbative exponential superpotentials generated, for instance, by 
gaugino condensation, and {\it added} to a perturbative superpotential are then incompatible with chiral-linear duality.\footnote{See however ref.~\cite{BDQQ1}
for an analysis.}

The effective four-dimensional supergravity depends on three moduli multiplets 
coupled to supergravity, gauge and charged matter superfields. Each of the three 
moduli scalars has a component of an antisymmetric tensor as supersymmetry 
partner. More precisely, the string coupling modulus 
is in the multiplet describing $G_{\mu\nu\rho y}$,
the $K_6$ volume modulus is paired with $G_{\mu i \ov k y}$ and the five-brane
$S^1/Z_2$ modulus is the partner of the components $\hat H_{\mu\nu\rho}$ 
or $\hat H_{\mu i \ov k}$ of the self-dual antisymmetric tensor living on the 
brane world-volume. We find, as explained in ref.~\cite{DSsugra}, more efficient
to formulate the theory using superconformal tensor calculus and to
introduce three moduli vector superfields to describe these moduli 
multiplets:\footnote{The Weyl weight is $w$ and $n$ is the chiral weight.}

\begin{tabular}{lll}
$V$ &($w=2, n=0$):  & $G_{\mu\nu\rho y}$, string coupling modulus, \ldots
\\
$V_T$ &($w=n=0$): & $G_{\mu i\ov k y}$, Calabi-Yau volume modulus, \ldots 
\\
$\hat V$ &($w=n=0$): & $\hat H_{\mu\nu\rho}$, five-brane $S^1/Z_2$ modulus,
\dots
\end{tabular}

\noindent
The components of the antisymmetric tensors are identified with the vector fields 
in these multiplets, the moduli scalars with their real lowest components.
These vector multiplets are then submitted to Bianchi identities
obtained from the $K_7$ reduction of the eleven-dimensional Bianchi identity
for $G_4$, eq. (\ref{bianchi1}), and the self-duality condition of the five-brane 
tensor. In each case, the supersymmetrized Bianchi identity also reduces the 
number of off-shell states from $8_B+8_F$ in a vector multiplet  
to $4_B+4_F$. These Bianchi identities are imposed as the field equations
of three Lagrange-multiplier superfields:

\begin{tabular}{lll}
$S$ &($w=n=0$):  & a chiral multiplet for the Bianchi identity verified by $V$,
\\
$L_T$ &($w=2, n=0$): & a linear multiplet for the Bianchi identity verified by
$V_T$,
\\
$\hat S$ &($w=n=0$): & a chiral multiplet for the self-duality condition of the 
\\ && brane tensor, applied to $\hat V$.
\end{tabular}

\noindent
Eliminating these Lagrange multiplier superfields defines the three vector multiplets 
in terms of the physical fields solving Bianchi identities. The important advantage of 
this procedure is that supersymmetrizing the theory before eliminating Lagrange 
multipliers automatically delivers the correct non-linear couplings of source terms 
(brane and plane contributions) to moduli and then the correct K\"ahler potential. 
Alternatively, equivalent (dual) versions of the theory can be obtained 
by eliminating some vector multiplets instead of the Lagrange multipliers. 

These six multiplets describing bulk and brane states are supplemented by states 
living on the fixed hyperplanes. In the notation defined in the previous section, these 
states include gauge chiral superfields ${\cal W}^{\alpha(i)}$ ($i=1,2$ as usual, $w=n=3/2$) 
and charged matter chiral multiplets ($w=n=0$) in some representation of the gauge group.
They will be collectively denoted by $M$ and they  contain the complex scalar components 
${\cal A}^{m(i)}$. Finally, we need
the compensating chiral multiplet $S_0$ ($w=n=1$) to gauge fix the superconformal
theory to super-Poincar\'e symmetry only. 

With this set of superfields, the Lagrangian nicely splits in a sum of
five terms with well-defined higher-dimensional interpretations:
\beq
\label{Leff1}
{\cal L} = {\cal L}_{bulk} +  {\cal L}_{Bianchi} +  {\cal L}_{kinetic} 
+  {\cal L}_{thresholds} + {\cal L}_{superpotential}.
\eeq
The bulk Lagrangian \cite{CFV}
\beq
\label{Lbulk}
{\cal L}_{bulk} = -  \Bigl[(S_0\ov S_0 V_T)^{3/2} (2V)^{-1/2} \Bigr]_D 
\eeq
can be directly obtained from the $K_7$ reduction of 
eleven-dimensional supergravity, expressed in terms of $G_4$. It depends 
on $V$ (string coupling multiplet) and $V_T$ ($K_6$ volume modulus multiplet),
and of the compensator $S_0$.  
In eq. (\ref{Lbulk}), $[\ldots]_D$ denotes the invariant real density formula of
superconformal calculus, as reviewed and developed in for instance refs.~\cite{KU}.
Similarly, $[\ldots]_F$ will below denote the chiral density formula.\footnote{In global supersymmetry, $[\ldots]_D$ and $[\ldots]_F$ would be $\int d^4\theta[\ldots]$ and 
$\int d^2\theta[\ldots] + {\rm h.c.}$}

The coupling of plane and brane fields (${\cal W}^{\alpha(i)}$, $M$ and $\hat V$) 
to bulk multiplets is entirely encoded in ${\cal L}_{Bianchi}$, which reads \cite{DS5brane}
\beq
\label{LBianchi}
{\cal L}_{Bianchi} = 
\Bigl[ - (S+\ov S)(V + 2\Omega_1 + 2\Omega_2 ) 
+ L_T (V_T + 2 \ov M e^{2A} M)
+ {1\over2} \tau (\hat S + \ov{\hat S}) V \hat V \Bigr]_D, 
\eeq
where $\Omega_1$ and $\Omega_2$ are the Chern-Simons multiplets ($w=2, n=0$) 
for the gauge algebra arising on each hyperplane, defined by\footnote{The operation 
$\Sigma(\ldots)$ is the superconformal analog of ${1\over8}\ov{\cal DD}$ in global superspace. }
$\sum_\alpha {\cal W}^{\alpha(i)} {\cal W}^{\alpha(i)} = 16\,\Sigma(\Omega_i)$, and
$\ov M e^{2A} M \equiv \sum_{m,i} \ov M^{m(i)} [e^{2A} M]^{m(i)}$ is the Wess-Zumino
Lagrangian superfield. This 
contribution is gauge-invariant since $[(S+\ov S)(\delta\Omega_1+\delta\Omega_2)]_D$ 
is a derivative. The dimensionless number $\tau$ is proportional to the five-brane tension 
$T_5$. In our units, it reads
\beq
\label{tauis}
\tau = {2\over(4\pi)^3} \, a,
\eeq
where $a$ is the area of the two-cycle wrapped by the brane  
in $K_6$, as defined in eq. (\ref{area}).
Notice shift symmetries $\delta S = ic$, $\delta\hat S= id$ ($c,d$ real).

The kinetic terms of the five-brane fields arise from reduction of the 
Dirac-Born-Infeld (DBI) Lagrangian:
\beq
\label{Lkinetic}
{\cal L}_{kinetic} = - \tau \bigl[ VV_T\hat V^2 \bigr]_D. 
\eeq
They are quadratic in $\hat V$, a consequence of the form of the DBI action, 
and the prefactor 
$VV_T$ is the coupling to the supergravity background. Notice that since this term
is linear in $V$, it will naturally assemble with the contribution in $S+\ov S$ in eq.
(\ref{LBianchi}). 

At this point, the contributions from bulk, plane and five-brane kinetic Lagrangians
have been considered, with tensor fields in $V$, $V_T$ and $\hat V$ verifying
Bianchi identities modified by plane and brane sources. But we still have to
consider further contributions from the topological term with modified $G_4$,
as obtained in the previous section. These terms will be collected in
${\cal L}_{threshold}$. The symmetries of expressions (\ref{Lbulk})--(\ref{Lkinetic}) 
allow the introduction of the following corrections to gauge kinetic terms
\cite{DS5brane} ({\it threshold corrections}):
\beq
\label{Lthreshold}
\begin{array}{rcl}
{\cal L}_{thresholds} &=& \Bigl[ -2\sum_i\beta_i\Omega_i (V_T+2\ov M e^{2A} M)\Bigr]_D
\crbig
&&
+ {\tau\over4}   \Bigl[\hat S \sum_i\hat\beta_i {\cal W}_i  {\cal W}_i \Bigr]_F
+  \Bigl[V \, \Bigl\{\epsilon|\alpha M^3|^2 - 2 \tau\sum_i g_i(\hat V) \ov M^i e^{2A} M^i 
\Bigr\} \Bigr]_D.
\end{array}
\eeq
The first contribution corresponds to threshold corrections depending on the
volume modulus \cite{DIN2, IN, DFKZ}. Gauge invariance of the full Lagrangian with
this term is obtained by postulating the appropriate variation of the linear 
multiplet $L_T$ in ${\cal L}_{Bianchi}$:
$$
\delta L_T = 2\sum_i\beta_i \delta \Omega_i.
$$
The second and third contributions are threshold corrections depending 
on the $S^1/Z_2$ location of the five-brane and/or, for the last one,
on matter multiplets. The coefficients $\beta_i, \hat\beta_i, \epsilon$ 
and the functions $g_i(\hat V)$  can be obtained, as explained below,
from Calabi-Yau reduction of the topological term with anomaly-cancelling 
modifications. Symmetries of the theory leave these functions unconstrained
but the terms considered here only require linear functions,  
$g_i(\hat V)=\hat \gamma_i \hat V +\hat \delta_i$.
Finally, the quantity $\alpha M^3$ in expression (\ref{Lthreshold})
denotes the holomorphic cubic invariant also present in the matter 
superpotential \cite{hetW, DIN2}.

The superpotential arises from the components $G_{ijk\, y}$ of $G_4$. They 
also verify a non-trivial Bianchi identity which is not modified
by five-branes since three holomorphic directions cannot be transverse to 
their world-volumes. The superpotential contribution to the Lagrangian
is then 
\beq
\label{Lsuperpotential}
{\cal L}_{superpotential} = \Bigl[ S_0^3\, W \Bigr]_F, 
\qquad{\rm with}\qquad
W = c + \alpha M^3,
\eeq
The constant $c$ being the `flux' of the heterotic three-form in direction 
$\epsilon_{ijk}$ \cite{hetW, gaugino}. 

This formulation of the effective supergravity, with six superfields to describe three
moduli supermultiplets leads to several equivalent forms, depending on the choice
made when eliminating the three superfluous multiplets. 
The Lagrange multipliers $S$, $L_T$ and $\hat S$ imply the following
constraints on the vector multiplets:
\beq
\label{Bianchisol}
\begin{array}{rrcll} 
S \, : & \quad V &=& L - 2\sum_i\Omega_i  , \quad &
L \,\, {\rm linear} \,\, (w=2, n=0),  
\crbig 
L_T \, : & \quad V_T &=& T + \ov T - 2 \ov M e^{2A} M, \quad &
T \,\, {\rm chiral}\,\, (w=n=0), 
\crbig
\hat S \, : & \quad \hat V &=& V^{-1}\left( \hat L  + 4 \sum_i\hat\beta_i\Omega_i\right), 
\quad & \hat L \,\, {\rm linear} \,\, 
(w=2, n=0).
\end{array}
\eeq
Eliminating $S$, $L_T$ and $\hat S$ leads then to a formulation where moduli
are described by $L$, $T$ and $\hat L$, two linear and one chiral multiplets:
\beq
\label{Llinear}
\begin{array}{rcl}
{\cal L} &=& \Bigl[ -{1\over\sqrt2} \, [ S_0\ov S_0
(T + \ov T - 2 \ov M e^{2A} M)]^{3/2} \,(L-2\sum_i\Omega_i)^{-1/2}
\crbig
&& -\tau (T + \ov T - 2 \ov M e^{2A} M) \, (L-2\sum_i\Omega_i)^{-1} \,
(\hat L + 4\sum_j\hat\beta_j\Omega_j)^2 
\crbig
&& + (L-2\sum_i\Omega_i) \, \Bigl\{\epsilon|\alpha M^3|^2 
- 2 \tau\sum_j \hat\delta_j  \ov M^j e^{2A} M^j \Bigr\}
\crbig
&& - 2\tau (\hat L  + 4 \sum_i\hat\beta_i\Omega_i) \sum_j \hat\gamma_j\ov M^j e^{2A} M^j 
\Bigr]_D
\crbig
&&+\Bigl [ S_0^3 W 
+ {1\over4}\sum_i  \beta_iT{\cal W}_i{\cal W}_i
\Bigr]_F .
\end{array}
\eeq
Component expressions for this apparently complicated theory can be
obtained using superconformal tensor calculus \cite{KU}. Notice that
plane contributions (superfields $\Omega_i$, ${\cal W}_i$ and $M^i$) now
appear in the bulk Lagrangian (first line) and also in the five-brane DBI 
contribution (second line). Five-brane contributions (multiplet $\hat L$)
appear in gauge kinetic terms (a ``plane term") while threshold corrections
(third and fourth lines) involve plane and five-brane fields. 
These mixings are induced by
the modified Bianchi identities, eqs. (\ref{Bianchisol}), and by threshold
corrections required by anomaly cancellation.

The kinetic term quadratic in the five-brane modulus superfield  $\hat L$
appears in the second line. It can clearly be derived from the DBI Lagrangian, as
done in ref. \cite{DS5brane}. But the superfield structure implied by the modified 
Bianchi identity leading to the first eq. (\ref{Bianchisol}) also 
implies that the same kinetic term can be obtained from 
gauge threshold corrections which follow from Calabi-Yau reduction of the (modified)
topological term. This point will be explicitly verified in the next paragraph. 

This unfamiliar supergravity theory is particularly useful to study its scalar potential and 
vacuum structure since linear multiplets do not have auxiliary fields. We will come 
back to this point later on and especially when studying condensation. 

It is however more common to formulate the supergravity theory with chiral 
moduli $S$ and $T$, and then to eliminate $V$ and $L_T$ using their field 
equations. One obtains:
\beq
{\cal L} = -{3\over2}\Bigr[ S_0\ov S_0 \, e^{-{\cal K}/3} \Bigl]_D 
+ \Bigr[ S_0^3 \, W  + {1\over4} \sum_i (S + \beta_iT + \tau \hat\beta_i \hat S)
\, {\cal W}^i {\cal W}^i \Bigl]_F ,
\eeq
with the K\"ahler potential
\beq
\label{Kis1}
{\cal K} = - \ln(S + \ov S - \Delta ) - 3 \ln(  T+\ov T - 2\ov M e^{2A} M )
\eeq
and
\beq
\label{Deltais}
\Delta = -\tau(T+\ov T - 2\ov M e^{2A} M) \hat V^2
+{1\over2}\tau (\hat S + \ov{\hat S}) \hat V +
\epsilon|\alpha M^3|^2 - 2\tau\sum_i g_i(\hat V) \ov M^i e^{2A} M^i  \, .
\eeq
The field equation of the vector multiplet $\hat V$ implies then
\beq
\label{Vrel}
\hat V =  (4V_T)^{-1}\,\Bigl( \hat S + \ov{\hat S} 
- 4\sum_i \hat \gamma_i \ov M^i e^{2A} M^i \Bigr),
\eeq
with $V_T$ as in the second eq. (\ref{Bianchisol}). The fully chiral formulation
of the effective supergravity is then defined by K\"ahler potential (\ref{Kis1})
with now
\beq
\label{Deltais2}
\begin{array}{rcl}
\Delta &=& 
{\tau\over16} (T+\ov T - 2\ov M e^{2A} M)^{-1}\Bigl( \hat S + \ov{\hat S}
- 4 \sum_i\hat\gamma_i \ov M^i e^{2A} M^i \Bigr)^2 
\crbig
&& + \epsilon|\alpha M^3|^2 - 2\tau\sum_i \hat\delta_i \ov M^i e^{2A} M^i  \, ,
\end{array}
\eeq
gauge kinetic functions
\beq
\label{kinf1}
f^i = S + \beta^i T + \tau \hat\beta^i\hat S,
\eeq
and superpotential (\ref{Lsuperpotential}). The presence of the five-brane then
introduces mixing of the three chiral multiplets $S$, $T$ and $\hat S$ and 
the off-diagonal
elements of the K\"ahler metric severely complicate the analysis of the theory.

Before returning to the analysis of the effective supergravity, we need a
precise identification of the supergravity fields in terms of massless modes
of the $K_7$ compactification. 

The notation we use for component fields is as follows. 
Vector multiplets $V$, $\hat V$ and $V_T$ have respectively vector fields
$V_\mu$, $\hat V_\mu$ and $T_\mu$ and (lowest) real scalar $C$, $\hat C$ and
$C_T$. And we use the same notation for chiral multiplets $S$, $T$ and $\hat S$ 
and for their complex scalar lowest components. Eqs. (\ref{Bianchisol})
indicate then that
$$
C_T = 2(\Re T - \ov MM), \qquad\qquad
T_\mu = -2\partial_\mu \Im T - 2i \ov M({\cal D}_\mu M)
+ 2i ( {\cal D}_\mu \ov M)M,
$$
that the lowest scalar component of the string coupling linear multiplet $L$ is also
$C$ and that the five-brane linear multiplet $\hat L$ has a real scalar 
$\hat\ell = C\hat C$. Relation (\ref{Vrel}) also implies that
\beq
\label{Vrel2}
\begin{array}{rcl}
\Re \hat S &=& 4 \hat C ( \Re T - \ov MM)
+ 2 \sum_i \hat \gamma_i \ov M^i  M^i\,,
\crbig
\partial_\mu \Im \hat S &=& - 4 \hat V_\mu (\Re T - \ov MM) 
+ 4\hat C (\partial_\mu \Im T + i\ov M{\cal D}_\mu M - iM{\cal D}_\mu\ov M)
\crbig
&&
- 2i \sum_i \hat \gamma_i (\ov M^i{\cal D}_\mu M^i - M^i{\cal D}_\mu\ov M^i).
\end{array}
\eeq
The scalar field $\hat C$ has background value proportional to $y$, the five-brane 
location along $S^1$. In our units with $2\pi R_{S^1}=1=V_6$, the
four-dimensional gravitational constant is $\kappa_4 = \kappa$ and
\beq
\label{hatCis}
\langle\hat C\rangle = {1\over\kappa_4} yR_{S^1} = {y\over 2\pi\kappa}.
\eeq 

In order to derive the identification of the matter scalars ${\cal A}^{m(i)}$ of
eq. (\ref{termI}) and the superconformal 
multiplets $M^{m(i)}$, we note that the component $G_{\mu i\bar{j}y}$ of the four-form 
is related to the vector component of the real multiplet $V_T$, which is
$$
(V_T)_\mu =
-2[\partial_\mu \Im T + i\ov M({\cal D}_\mu M) - i ({\cal D}_\mu \ov M) M].
$$
On the other hand, using eqs. (\ref{termB}) and (\ref{termI}),
we also find
$$
\begin{array}{rcl}
G_{\mu i\bar{j}y}&=&\displaystyle{
\partial_{\mu}C_{i\ov j y} -\frac{(4\pi\kappa^2)^{1/3}}{\pi}\,
\sum_k \omega_{k, \mu i \ov j}}
\\ 
&=&\displaystyle{
\frac{1}{2\pi}\,(4\pi\kappa^2)^{-2/3}\,
\Big( \partial_{\mu} \Im T - i {\kappa^2\over2\pi} 
\big[ {\cal A}^{(i)}{\cal D}_{\mu}\ov {\cal A}^{(i)} -
\ov{\cal A}^{(i)}{\cal D}_{\mu} {\cal A}^{(i)}\,\Big]
\Big)\,\delta_{i\bar{j}}\,.}
\end{array}
$$
By comparison, one obtains
\beq
{\cal A}^{(i)} \,=\,\frac{\sqrt{2\pi}}{\kappa}\,M^{(i)} ,
\eeq
with an irrelevant sign choice.

Finally, the gauge fields in $\Omega_i$ or ${\cal W}^{a(i)}$ are
the massless modes ${\cal A}_\mu^{a(i)}$.

\subsection{Gauge coupling constants}\label{secgsquare}

The effective supergravity Lagrangian predicts a very specific moduli dependence 
in the super-Yang-Mills sector relevant to the determination of the effective
superpotential with non-perturbative configurations. 
The field-dependent gauge couplings can of course 
be obtained in any formulation of the theory. But the closest relation to the 
higher-dimensional origin of the effective supergravity
is realized with supermultiplets $L$ (for the string dilaton multiplet), $T$ (Calabi-Yau
volume modulus) and $\hat L$ (for the five-brane $S^1/Z_2$ modulus). This is the
theory defined by eq. (\ref{Llinear}) in which tensor calculus leads to the following
(inverse squared) gauge couplings:
\beq
\label{glinear}
\begin{array}{rcl}
\displaystyle{1\over g_i^2} &=& \displaystyle
{1\over2} \left({z_0\ov z_0 (T+\ov T - 2\ov MM) \over 2C}\right)^{3/2}
+ {\tau\over2}(T+\ov T - 2\ov MM)[ \hat C^2 + 4\hat\beta_i \hat C ]
\crbig
&&
+ \beta_i \Re T 
+{1\over2}\left[ \epsilon |\alpha M^3|^2 - 2\tau \sum_j (\hat\delta_j
-\hat\beta_i\hat\gamma_j) \ov M^j M^j \right]. 
\end{array}
\eeq
The complex scalar $z_0$ is the lowest component of the compensating multiplet
$S_0$. In the Poincar\'e theory, it is a function of the physical scalars chosen to
obtain a specific ``gravity frame". The Einstein frame where the gravity Lagrangian
is $-{1\over2\kappa_4^2}e_4R_4$ corresponds to\footnote{See for instance 
\cite{DQQ, DS5brane}.}
\beq
\label{z0is}
2\kappa_4^2 C = \left({z_0\ov z_0 (T+\ov T - 2\ov MM) \over 2C}\right)^{-3/2}.
\eeq
Without branes or threshold corrections the dimensionless field $4\kappa_4^2 C$
is then the (universal) gauge coupling $g_i^2$.

On the other hand, the chiral version of the theory has gauge kinetic functions 
(\ref{kinf1}) and then:
\beq
\label{gchiral}
{1\over g_i^2} = \Re (S + \beta_i T + \tau \hat\beta_i \hat S).
\eeq
The equality of these two expressions is encoded in the duality
transformations exchanging linear multiplets $L$ and $\hat L$ with $S$ and $\hat S$:
\beq
\label{dualtransf}
\begin{array}{rcl}
{1\over2}( \hat S + \ov{\hat S}) &=& 2(T+\ov T - 2\ov Me^{2A}M) \hat V
+ 2\sum_i \hat\gamma_i \ov M^i e^{2A} M^i , 
\crbig
{1\over2}( S + \ov S) &=& {1\over2} \left({ S_0\ov S_0 (T+\ov T - 2\ov Me^{2A}M)
\over 2 V }\right)^{3/2}
+ {1\over2} \left[ \epsilon|\alpha M^3|^2 - 2\tau\sum_i\hat\delta_i\ov M^i e^{2A} M^i  \right]
\crbig
&& + {\tau\over2} (T+\ov T - 2\ov Me^{2A}M) \hat V^2,
\end{array}
\eeq
with $V$ and $\hat V$ as in eqs. (\ref{Bianchisol}). The lowest scalar components of these 
two superfield equations show the equality of (\ref{glinear}) and (\ref{gchiral}).

The ``natural" definition of the dilaton $\varphi$ with kinetic Lagrangian 
$(\partial_\mu\ln\varphi)^2/4$ is to identify
$$
-\ln( S + \ov S - \Delta) \quad\longleftrightarrow\quad - \ln (2\varphi)
$$
in the K\"ahler potential (\ref{Kis1}), {\it i.e.}
\beq
\label{dilaton}
\begin{array}{rcl}
\varphi &=&  \displaystyle
{1\over2} \left({z_0\ov z_0 (T+\ov T - 2\ov MM) \over 2C}\right)^{3/2}
\crbig
&=& \displaystyle
\Re S - {\tau\over32}  {[\hat S +\ov{\hat S} - 4\sum_i \hat\gamma_i \ov M^i M^i ]^2
\over T+\ov T-2\ov MM }
+\tau\sum_i \hat\delta_i \ov M^i M^i
-\frac{1}{2}\epsilon |\alpha
M^{3}|^{2}\,.
\end{array}
\eeq

From now on, we omit charged matter terms, as we expect $\langle M\rangle=0$.
In terms of the dilaton, the gauge couplings read then
\beq
\label{g2}
\begin{array}{rcl}
\displaystyle{1\over g_i^2} &=& \varphi + \Re T \left[ \tau \hat C^2  
+ 4\tau\hat\beta_i\hat C + \beta_i \right]
\\
&=& \displaystyle \varphi + \Re T \Bigl[ {\tau\over16} \left({\Re\hat S\over\Re T}\right)^2 
+ \tau\hat\beta_i{\Re\hat S\over\Re T} + \beta_i \Bigr].
\end{array}
\eeq
They display a universal\footnote{{\it i.e.}~identical 
for all group factors, all values of index $i$.} correction quadratic in the five-brane location, 
as well as gauge group-dependent 
corrections linear in $\hat C$ or constant. The chiral version has only terms linear 
in $S$, $T$ and $\hat S$: the universal quadratic correction has been absorbed in the 
non-harmonic redefinition turning the dilaton $\varphi$ into $\Re S$. And of course
the quadratic term reappears in the K\"ahler potential for $S$ [see eqs. (\ref{Kis1}) 
and (\ref{Deltais2})]. 

We now restore the summation over several five-branes and split the coefficients $\beta_i$
according to $\beta_i = \beta_i^{(pl.)} +  \beta_i^{(br.)}\sum_I\tau_I $ since they receive in general 
contributions from both planes and five-branes.\footnote{The constants $\beta_i^{(pl.)}$,
$\beta_i^{(br.)}$ and $\hat\beta_i$ should not depend on $I$.} Using the identification 
(\ref{hatCis}), one obtains 
\beq
\label{gyis}
{1 \over g_i^2}= \varphi + \beta_i^{(pl.)} \Re T + \Re T \sum_I \tau_I 
\Bigl[ \beta^{(br.)}_i + \frac{4}{\kappa}\hat{\beta}_{i}\left( \frac{y_{I}}{2\pi }\right)
+\frac{1}{\kappa^2}\left( \frac{y_{I}}{2\pi }\right) ^{2}\Bigr] .
\end{equation}
Notice that since the exchange $1\leftrightarrow2$ of the plane indices is equivalent 
to moving the five-brane from $y_I$ to $\pi- y_I$, we expect 
$$
\hat\beta_1 + \hat\beta_2 = - {1\over4\kappa},
\qquad\qquad
\beta_2^{(br.)} = \beta_1^{(br.)} + {2\over\kappa} \hat\beta_1
+{1 \over 4\kappa^2}.
$$

The next step is to compare these results with the terms obtained from the reduction 
of the topological term and especially with the brane contributions described by the
first term in eq. (\ref{Lbrane}):
\beq
\begin{array}{rcll}
\label{Lbrane22}
e_{4}^{-1}\,\Delta \mathcal{L}_{brane} &=& \displaystyle
{1\over4} \sum_{I}\frac{\tau _{I}}{\kappa ^{2}} \,\mbox{Im}T
\Biggl\{ \left[ \left( \frac{y_{I}}{2\pi }\right) ^{2}-\frac{y_{I} 
}{2\pi }+\frac{1}{6}\right] \mbox{Tr} \, F^{(1)}\widetilde{F}^{(1)}
\crbig && \hspace{1.9cm}\displaystyle
+\left[\left( \frac{y_{I}}{2\pi }\right) ^{2}-\frac{1}{12}\right] \mbox{Tr} \, F^{(2)} 
\widetilde{F}^{(2)}\Biggr\} \,.
\end{array}
\eeq
The terms of order $(y_I)^2$ have identical coefficient for both planes. 
If our effective supergravity is correct, this contribution should appear in the 
DBI term appearing in the second line of Lagrangian (\ref{Llinear}).
The vector component of the Chern-Simons multiplet $\Omega$ is
$$
\lbrack \Omega ]_{vector} \equiv \Omega^\mu =
\frac{1}{8e_{4}}\varepsilon ^{\mu \nu \rho \sigma
}\omega _{\nu \rho \sigma }+... 
$$
and the component expansion of (\ref{Llinear}) includes then
$$
- 4 e_4 \,\tau_{I}\, \hat{C}_{I} ^{2}\,(\partial _{\mu } \Im T)
 \,\sum_{i}\Omega _{i}^{\mu }+ \ldots
$$
Integrating by part for constant values
$\hat C_I = \langle \hat C_I\rangle = y_I/(2\pi\kappa)$, we obtain
\beq
\label{Lbrane2bis}
 \displaystyle  {1\over4} e_4 \frac{\tau _{I}}{\kappa ^{2}}%
\left( \frac{y_{I}}{2\pi }\right) ^{2}\,\mbox{Im}T\,\sum_{i} \Tr \,
F^{(i)}_{\mu \nu }\widetilde{F}^{(i)\mu \nu }+ \ldots
\eeq
which fits correctly the quadratic term in (\ref{Lbrane22}). It is then not 
needed to perform the Calabi-Yau reduction of the five-brane DBI Lagrangian
to find the kinetic terms of the five-brane modulus superfield: knowledge of the
superfield structure, eq.~(\ref{Llinear}), and of the gauge terms implied by 
the topological term is sufficient.
Similarly, the terms of order $y_I$ and $(y_I)^0$ in the effective supergravity and in the 
reduction of the topological term can be used to find the values of the 
coefficients $\beta _{i}^{(2)}$ and $\hat{\beta}_{i}$.

The second line of Lagrangian (\ref{Llinear}) indicates that the DBI contribution
also includes gauge terms of order $\hat C$, which after partial integration read
\beq
\frac{1} {\kappa }   \Im T
\sum_I \tau_I \, \frac{y_{I}}{2\pi } \,
\sum_{i}\hat{\beta}_{i}\Tr\, F^{(i)}_{\mu \nu } \widetilde{F}^{(i)\mu \nu }. 
\eeq
By comparison with (\ref{Lbrane22}), we find
\beq
\label{hatbetaare}
\hat{\beta}_1 = - \frac{1}{4\kappa},
\qquad\qquad
\hat{\beta}_2 = 0.
\eeq
Finally, comparison of the $y_I$--independent terms in eq. (\ref{Lbrane22})
with the first term of (\ref{Lthreshold}), which includes
\beq
-2\sum_{i}\left[ \beta_{i}\Omega _{i}V_{T}\right] _{D}= \frac{1}{4} \Im T
\,\sum_{i}\beta_{i} \Tr F^{(i)}_{\mu \nu }\widetilde{F}^{(i)\mu \nu }+ \ldots,
\eeq
indicates that
\beq
\label{2beta}
\beta_1^{(br.)} = \frac{1}{6\kappa^2} \,,
\qquad\qquad
\beta_2^{(br.)} =  - \frac{1}{12\kappa^2} \,.
\eeq
As expected, exchanging planes $1\leftrightarrow2$ is equivalent to 
$y_I \leftrightarrow \pi - y_I$. 

Finally, as usual, the coefficients $\beta_i^{(pl.)}$ can be read directly from the
first line of eq. (\ref{Lplane}), which includes contributions to the topological terms
arising from the hyperplanes only:
\beq
\beta_i^{(pl.)} = {2\over (4\pi)^4\kappa^2}\sum_j d_{ij} \, \langle{\cal I}_j\rangle, 
\eeq
or
\beq
\beta_1^{(pl.)} = {1\over 6(4\pi)^3\kappa^2}\Bigl(2 \langle{\cal I}_1\rangle
- \langle{\cal I}_2\rangle \Bigr), \qquad\qquad
\beta_2^{(pl.)} = {1\over 6(4\pi)^3\kappa^2}\Bigl(2 \langle{\cal I}_2\rangle
- \langle{\cal I}_1\rangle \Bigr). 
\eeq
Notice that 
$$
\beta_1^{(pl.)} +\beta_2^{(pl.)} = {1\over 6(4\pi)^3\kappa^2}\Bigl( \langle{\cal I}_1\rangle
+ \langle{\cal I}_2\rangle \Bigr)
$$
vanishes in the minimal embedding without five-brane \cite{DIN2, IN}. 

To summarize, in terms of the dilaton, the gauge couplings on both planes, as 
calculated from the modified topological term, read:
\beq
\begin{array}{rcl}
\displaystyle{1\over g_1^2} &=& \displaystyle \varphi + \beta_1^{(pl.)} \Re T
+{1\over\kappa^2} \Re T \sum_I \tau_I \left[ \left( {y_I\over2\pi} \right)^2
- \left( {y_I\over2\pi} \right) + {1\over6} \right],
\crbig
\displaystyle{1\over g_2^2} &=& \displaystyle \varphi + \beta_2^{(pl.)} \Re T
+{1\over\kappa^2} \Re T \sum_I \tau_I \left[ \left( {y_I\over2\pi} \right)^2
- {1\over12} \right].
\end{array}
\eeq
A nicer expression reminiscent of eqs. (\ref{topol4}) or (\ref{12form})
uses the distance from the brane to the planes:
\beq
\label{g2fin}
\begin{array}{rcl}
\displaystyle{1\over g_1^2} &=& \displaystyle \varphi + \beta_1^{(pl.)} \Re T
+{1\over 4\kappa^2} \Re T \sum_I \tau_I \left[ (\Delta_{I,2})^2 - {1\over3} \right],
\crbig
\displaystyle{1\over g_2^2} &=& \displaystyle \varphi + \beta_2^{(pl.)} \Re T
+{1\over 4\kappa^2} \Re T \sum_I \tau_I \left[(\Delta_{I,1})^2 - {1\over3} \right].
\end{array}
\eeq
where $\Delta_{I,1} = y_I/\pi$, and $\Delta_{I,2} = 1 - y_I/\pi$, as in eq. (\ref{Deltadef}).
The contribution of a five-brane to the gauge couplings on one hyperplane decreases
quadratically from a maximum value
$$
{1\over 6\kappa^2} \tau_I \Re T
$$
when the brane lies on the plane, to a minimal value
$$
-{1\over 12\kappa^2} \tau_I \Re T
$$
when the five-brane lies on the opposite plane. For a five-brane in the middle 
of the interval, both gauge couplings receive the correction
$$
-{1\over 48 \kappa^2} \tau_I \Re T.
$$
With however $(\Delta_{I,i})^2 = - \Delta_{I,1}\Delta_{I,2} + \Delta_{I,i}$, ($i=1,2$),
the term quadratic in $y$ is necessarily universal and the two (inverse squared) 
gauge couplings differ only by a contribution linear in $y$:
\beq
\label{g2diff}
{1\over g_1^2} - {1\over g_2^2} = {\Re T \over 2(4\pi)^3\kappa^2}
\left[  \langle{\cal I}_1\rangle -  \langle{\cal I}_2\rangle 
+ \sum_I a_I \Bigl( 1 - {2y_I\over\pi} \Bigr) \right],
\eeq
with $\Delta_{I,2} - \Delta_{I,1} = 1 - 2y_I/\pi$ and
in terms of the area $a_I$ of the two-cycle wrapped by the five-brane in $K_6$
[see eq. (\ref{tauis})]. 

The normalization of the four-forms $I_{4,1}$ and $I_{4,2}$ is such that their integrals
over a four-cycle in $K_6$  are half-integers. Similary $a_I$ is an intersection 
number of the two-cycle wrapped by the brane with the four-cycle, in units of the 
Calabi-Yau volume.\footnote{We have chosen $V_6=1$.} These statements follow
from the integrated Bianchi identity verified by $G_4$
and from rewriting eqs. (\ref{Iiis}) and (\ref{area}) in the form 
$$
a_I  = -i \int_{K_6} \delta^{(4)}({\cal W}_{6,I}) \wedge dz \wedge d\ov z,
\qquad\qquad
\langle {\cal I}_i \rangle = -i \int_{K_6} \langle I_i\rangle \wedge dz\ d\ov z. 
$$
Eq. (\ref{g2diff}) matches then nicely the idea that a five-brane moved to the 
hyperplane at $y=0$ decreases the instanton number on this plane, or on the
second plane when moved to $y=\pi$. 

In the chiral version of the theory, the gauge kinetic functions are
\beq
\label{f1f2}
\begin{array}{rcl}
f_2 &=& \displaystyle S + \beta_2^{(pl.)} T - {1\over 12 \kappa^2} \sum_I \tau_I T , 
\crbig
f_1 &=& \displaystyle f_2 + \left[\beta_1^{(pl.)} - \beta_2^{(pl.)} \right] T 
+  {1\over 4 \kappa^2} \sum_I \tau_I \left[ T -  \kappa  \hat S \right]. 
\end{array}
\eeq
Since
$$
\langle\Re T\rangle -  \kappa  \langle\Re \hat S \rangle
=  \langle\Re T\rangle \Bigl(1 - {2y\over\pi} \Bigr),
$$
the difference is again eq. (\ref{g2diff}).

\subsection{Discussion of some matter terms}

We have seen that the gauge part of the five-brane-induced contributions to the topological 
terms are due, in the effective supergravity, either to the effect of the modified Bianchi identities
on the Dirac-Born-Infeld Lagrangian, or to threshold corrections. 
Since charged matter multiplets $M$ arise
from the fields living on the fixed hyperplanes, as do gauge multiplets, we may
expect that some or all matter contributions obtained from the reduction of the
topological term can also be derived from the DBI effective Lagrangian. 

As an illustration, we will establish that the charged matter term in expression 
(\ref{Lbrane}) arises for its part quadratic in $y_I$ from the DBI Lagrangian, while 
the terms linear and constant in $y_I$ originate from threshold corrections.
Since the vector component of $V$ is related to the string antisymmetric
tensor $B_{AB}$ by
\beq
v^{\mu} =  -\frac{2\pi}{8e_4}\epsilon^{\mu\nu\rho\sigma}
\partial_{\nu}B_{\rho\sigma}\,,
\eeq
the string--matter coupling term in eq. (\ref{Lbrane}) takes the form
\beq
\label{Matterm}
\begin{array}{l}
 \displaystyle{
- \frac{i}{8(4\pi )^3} \sum_{I,i}  a_I
\left( \Delta _{I,i}^2 - {1\over3} \right)
 \, \epsilon^{\mu \nu \rho \sigma }\partial_\mu B_{\nu \rho}
\sum_m [ {\cal A}^{m(i)}(\mathcal{D}_{\sigma }\ov{\cal A}^{m(i)})\,-\,
\ov{\cal A}^{m(i)}(\mathcal{D}_{\sigma }{\cal A}^{m(i)})]  } 
\crbig \hspace{2cm}
=\displaystyle{ 
- \frac{2 i }{\kappa^2} e_4\sum_{I}  \tau_I
v^{\mu}\Big\{ \left({y_I\over2\pi}\right)^2\,
 [ M(\mathcal{D}_{\mu }\ov M)\,-\,
\ov M(\mathcal{D}_{\mu } M)]  \qquad\qquad } 
\crbig \hspace{4.5cm}
\displaystyle{ \,+ \,\sum_i \tilde{g}_i(y_I)
[ M^i(\mathcal{D}_{\mu }\ov M^i)\,-\,
\ov M^i(\mathcal{D}_{\mu } M^i)] 
\Big\}\, . }
\end{array}
\eeq
The functions $\tilde{g}_i(y_I)$ are at most linear in $y_I$:
\beq
\tilde{g}_1(y_I) =  -\frac{y_I}{2\pi}+\frac{1}{6}
\,, \qquad\qquad \tilde{g}_2(y_I) = - \frac{1}{12}\,.
\eeq
The first term in the {\it r.h.s.} of eq. ({\ref{Matterm}}) is universal  and can clearly be 
retrieved from the DBI Lagrangian in the second line of theory (\ref{Llinear}) 
by selecting the matter contribution inside $V_T$,
\beq
- {2i e_4  \tau_I \,\hat C_I^2\, v^{\mu} 
[ M(\mathcal{D}_{\mu }\ov M)\,-\,
\ov M(\mathcal{D}_{\mu } M)]\,+ \ldots \,,}
\eeq
and identifying as usual $\hat C^2$ with $y_I^2 (2\pi\kappa)^{-2}$. 

The second term in the {\it r.h.s.} of eq. ({\ref{Matterm}}) originates from matter 
threshold corrections
in the supergravity Lagrangian (\ref{Llinear}). The relevant term is:
\beq
-2ie_4\tau_I v^{\mu} 
 \sum_i g_i( \hat C)[ M^i(\mathcal{D}_{\mu }\ov M^i)\,-\,
\ov M^i(\mathcal{D}_{\mu } M^i)]+ \ldots
\eeq
Comparison with eq. (\ref{Matterm}) indicates that
\beq
\hat\gamma_1 = - \frac{1}{\kappa} 
\qquad
\hat\gamma_2 = 0\,,
\qquad
\hat\delta_1 = \frac{1}{6\kappa^2} \,, 
\qquad 
\hat\delta_2 = -\frac{1}{12\kappa^2}\,.
\eeq
Interestingly enough, $\hat\delta_i = \beta_i^{(br.)}$ and $\hat\gamma_i = 4\hat\beta_i$. 

Finally, we briefly return to the issue of tracing back the supersymmetric
origin of interactions such as expressions (\ref{Lplane2}) and (\ref{Lbrane2}). As already
mentioned in Section 3, eq. (\ref{Lplane2}) seems in general hard to 
cast in a supersymmetric form because of the complicated mixing of matter contributions
from both hyperplanes. This feature is also present
in eq. (\ref{Lbrane2}), forbidding by the same token its supersymmetrization for a generic 
background. We however have little information on the nature of four-dimensional 
matter counter-terms which could be added to anomaly-cancelling corrections and
could radically change the picture. At this level of understanding, this discussion cannot
be conclusive.

Nevertheless, eq. (\ref{Lplane2}) allows a supersymmetric formulation for the standard 
embedding in the gauge group in which chiral matter multiplets only appear on one 
plane, say ${\cal A}\equiv {\cal A}_2$. Then,  eq. (\ref{Lplane2}) reduces to 
\beq
\label{Lplane22}
\begin{array}{rcl}
\Delta \mathcal{L}_{plane}  &=&  {\ds
i \,\epsilon^{\mu \nu \rho \sigma}\,\omega_{1,\mu \nu \rho}\,\Big\{
{1\over 3(4\pi)^5}\,[
\lambda{\cal A}^3\; \partial_{\sigma}(\overline{\lambda{\cal A}}^3)
-\overline{\lambda{\cal A}}^3\; \partial_{\sigma}(\lambda{\cal A}^3)]\,}
\crbig
&&
- {\ds {1 \over 12(4\pi)^3}
\sum_{m}({\cal A}^{m}\mathcal{D}_{\sigma }\ov {\cal A}^{m}-
\ov  {\cal A}^{m}\mathcal{D}_{\sigma}{\cal A}^{m})\, (\langle {\cal I}_1\rangle - \langle {\cal I}_2\rangle)\,\Big\}\,,}
\end{array}
\eeq 
which extends to the supersymmetric density
\beq
{1 \over 3(4\pi)^2\kappa^6}\,\left[\Omega_{1}\,|\alpha M^3|^2\right]_D 
- {1 \over 3(4\pi)^2\kappa^2}\,(\langle {\cal I}_1\rangle - \langle {\cal I}_2\rangle) 
\left[\Omega_{1}\,\ov M e^{2A} M\right]_D.
\eeq
The second term could in principle correspond to the first contribution appearing
in the threshold correction (\ref{Lthreshold}).

\section{Condensation, the non-perturbative superpotential and membrane 
instantons}\label{seccond}
\setcounter{equation}{0}

The non-perturbative superpotential arises from the $F$--density in the 
supergravity Lagrangian (\ref{Leff1}) when some or all gauge fields condense. 
It can be evaluated using a standard two-step procedure: first obtain
the effective action for condensates and then eliminate the condensate by solving 
(approximately in general) its field equation. Computing the effective action amounts in principle
to couple the superfield ${\cal WW}$ to an external source, integrate the gauge fields
and perform the Legendre transformation exchanging 
the source field with the (classical) condensate field. It is well known that
the symmetry content of super-Yang-Mills theory and anomaly-matching are 
sufficiently restrictive to accurately describe condensation \cite{VY}.  

As usual, we assume that the gauge multiplet which first condenses does not 
couple to matter multiplets $M$. We then split 
the gauge group into ${\cal G}_0 \times \prod_a {\cal G}_a$, where the simple
group ${\cal G}_0$ condenses and matter multiplets only tranform under 
$\prod_a {\cal G}_a$. The terms involving ${\cal G}_0$ gauge fields in the 
Lagrangian reduce then to 
\beq
\label{gaugeonly}
{1\over4} \left[ (S + \beta_0 T + \tau \hat\beta_0 \hat S) {\cal W}_0{\cal W}_0 
\right]_F. 
\eeq
Following ref.~\cite{VY}, these contributions are simply replaced in the effective 
action for condensates by
\beq
\label{Lcond1}
 {1\over4}  \left[(S + \beta_0 T + \tau \hat\beta_0 \hat S) U + 
{b_0\over24\pi^2}\left\{ U\ln \left( { U \over \mu^3} \right) - U\right\}  \right]_F
+ \left[ S_0\ov S_0 \, \widetilde K \left( US_0^{-3}, \ov U\ov S_0^{-3} \right) 
\right]_D ,
\eeq
where $U$ is the (classical) chiral superfield ($w=n=3$) describing the gaugino
condensate, $\langle U \rangle = \langle\lambda\lambda\rangle$. 
The coefficient of the Veneziano-Yankielowicz superpotential
is dictated by the anomaly of the superconformal chiral $U(1)$. It is proportional to 
the one-loop beta function, $b_0 = 3C({\cal G}_0)$, and the scale parameter $\mu$ 
is the energy at which gauge couplings in expressions (\ref{g2fin}) are defined. 
Finally, the leading contribution to the K\"ahler potential $\widetilde K$ is 
controlled by the scaling dimension (Weyl weight) of $U$.

The effective Lagrangian with condensate $U$ can be obtained by collecting all 
terms in the ``microscopic" Lagrangian (\ref{Leff1}), with contributions
(\ref{gaugeonly}) replaced by the effective terms (\ref{Lcond1}):
\beq
\label{Lcond2}
\begin{array}{rcl}
{\cal L}_{eff.} &=& \Bigr[
- \{ S_0\ov S_0 (T + \ov T - 2\ov M e^{2A} M)\}^{3/2} (2V)^{-1/2}
- (S+\ov S)V 
\crbig
&&
- \tau V\hat V^2 (T + \ov T - 2\ov M e^{2A} M) +{\tau\over2}(\hat S + \ov{\hat S})
V\hat V
\crbig
&& + V  \{\epsilon|\alpha M^3|^2 - 2\tau \sum_i g_i(\hat V) \ov M^i e^{2A} M^i \} 
+ S_0\ov S_0 \, \widetilde K \left( US_0^{-3}, \ov U\ov S_0^{-3} \right) \Bigr]_D
\crbig
&& + {1\over4}   \Bigl[ \sum_a( S + \beta_a T+ \tau \hat\beta_a \hat S) 
{\cal W}_a  {\cal W}_a \Bigr]_F
\crbig
&& + \Bigl[ S_0^3\, (c + \alpha M^3) 
+ {1\over4} (S + \beta_0 T + \tau \hat\beta_0 \hat S) U 
+ {C({\cal G}_0)\over32\pi^2} \{ U\ln(U/\mu^3) - U \} \Bigr]_F. 
\end{array}
\eeq
As before, vector multiplets $V$ and $\hat V$ are constrained by the field equations
of Lagrange multipliers $S$ and $\hat S$, which impose modified Bianchi identities. 
Rewriting their contributions in the form
\beq
\label{Bianchinew}
\Bigl[ 2S \Bigl\{ \Sigma(V) + {1\over8} U + {1\over8}\sum_a {\cal W}_a{\cal W}_a \Bigr\} 
- \tau \hat S \Bigl\{ \Sigma(V\hat V) - {1\over4} \hat\beta_0 U - 
{1\over4}\sum_a \hat\beta_a{\cal W}_a{\cal W}_a \Bigr\} \Bigr]_F,
\eeq
multiplier $S$ implies
\beq
\label{Simplies}
U = -8\Sigma(V_0), \qquad\qquad V = V_0 - 2\sum_a\Omega_a,
\eeq
with a real vector multiplet $V_0$ (8 bosons $+$ 8 fermions). Then, multiplier
$\hat S$ requires
\beq
\label{hatSimplies}
V\hat V = \hat L_0 - 2\hat\beta_0V_0 + 4\sum_a\hat\beta_a\Omega_a, 
\eeq
with a linear multiplet $\hat L_0$ and $V$ as given in eqs. (\ref{Simplies}).
These solutions can be compared with eqs. (\ref{Bianchisol}), which apply
before condensation of ${\cal W}_0{\cal W}_0$. Clearly, the real vector multiplet
$V_0$ describes $L - 2\Omega_0|_{cond.}$, {\it i.e.}~it includes the string 
coupling linear multiplet $L$ and the condensate field \cite{BDQQ1}. Similarly, the
linear multiplet $\hat L_0$ replaces $\hat L + 2\hat\beta_0 L$. 

To obtain the non-perturbative superpotential however, one first chooses the 
formulation of the theory with chiral multiplets only. The elimination of vector 
multiplets $V$ and $\hat V$ is as in the previous section. 
Omitting from here on gauge fields ${\cal W}_a$ and charged 
matter fields $M$, the chiral formulation of the effective Lagrangian is
\beq
\label{Leff2}
\begin{array}{rcl}
{\cal L}_{eff.} &=& \Bigl[ -{3\over2}S_0\ov S_0 \, e^{-{\cal K}/3} + 
S_0 \ov S_0 \widetilde K \Bigr]_D
\crbig
&& + \Bigl[ S_0^3\, (c + \alpha M^3) 
+ {1\over4} (S + \beta_0 T + \tau \hat\beta_0 \hat S) U 
+ {C({\cal G}_0)\over32\pi^2} \{ U\ln(U/\mu^3) - U \} \Bigr]_F .
\end{array}
\eeq
The K\"ahler potential ${\cal K}$ is as in eqs. (\ref{Kis1}) and (\ref{Deltais2}),
with full mixing of multiplets $S$, $T$ and $\hat S$. 
To derive the non-perturbative effective potential, neglect 
$\widetilde K$.\footnote{Disregarding the K\"ahler 
potential $\widetilde K$ is the same as considering $U$ as a constant 
background field with value chosen to extremize the action.} The
field equation of the condensate field $U$ implies then
\beq
\label{Uis}
U = \mu^3 \, {\rm exp}\left(-{8\pi^2 \over  C({\cal G}_0)} 
[ S + \beta_0 T + \tau \hat\beta_0 \hat S ] \right)
\equiv  {\cal U} ,
\eeq
and the effective superpotential becomes 
\beq
\label{Wnpis}
W_{np} = c + \alpha M^3
-  {C({\cal G}_0)\over32\pi^2}  \, {\cal U} \, S_0^{-3} . 
\eeq
This superpotential is the {\it sum} of the `microscopic' superpotential and
the non-pertur\-ba\-ti\-ve contribution of the gauge condensate. The 
non-perturbative contribution is the {\it exponential of the sum} of the string coupling, 
Calabi-Yau volume and five-brane moduli contributions. 
With several condensates, the 
non-perturbative piece would be replaced by a {\it sum of similar terms} 
over all gauge simple factors which condense. 

To get a qualitative picture of the effect of the five-brane, use eqs. (\ref{f1f2}) 
and (\ref{g2diff})
to rewrite the condensate as a function of the five-brane position along $S^1$,
assuming first that the condensate arises on the hyperplane at $y=0$:
\beq
\label{condy=0}
\begin{array}{rcl}
|U|_{1st \,\, plane}
&=& 
|\mu|^3 \, {\rm exp}\left(-{8\pi^2 \over  C({\cal G}_0)} 
\Bigl[ \Re S + {1\over (4\pi)^3\kappa^2} \Re T \Bigl \{ 
{1\over3}\langle {\cal I}_1\rangle  - {1\over6}\langle {\cal I}_2 \rangle
+{1\over3}a -{y\over\pi} a \Bigr\} \Bigr] \right)
\crbig
&=& |\mu|^3 \, {\rm exp}\left(-{8\pi^2 \over  C({\cal G}_0) g_1^2} \right) .
\end{array}
\eeq 
If the condensate arises on the hyperplane at $y=\pi$:
\beq
\label{condy=pi}
\begin{array}{rcl}
|U|_{2nd \,\, plane} 
&=& |\mu|^3 \, {\rm exp}\left(-{8\pi^2 \over  C({\cal G}_0) g_2^2} \right)
\crbig
&=& |U|_{1st \,\, plane} \times
{\rm exp}\left( {4\pi^2 \over C({\cal G}_0)} {\Re T \over (4\pi)^3\kappa^2}
\Bigl[ \langle {\cal I}_1\rangle - \langle {\cal I}_2\rangle 
+ a (1 - {2y\over\pi})\Bigr] \right).
\crbig
\end{array}
\eeq 

The non-perturbative superpotential (\ref{Wnpis}) and the condensates 
(\ref{condy=0}) and (\ref{condy=pi}) display the dependence
on the five-brane location on $S^1$ expected from explicit calculations of membrane 
instanton corrections in the four-dimensional effective theory \cite{MPS,
2brinst}. We have obtained this dependence from the analysis of the fundamental
Bianchi identity of M-theory and from the (modified) topological term,
showing in this way that open membrane instanton contributions find their higher-dimensional
origin in anomaly-cancellation in the presence of five-branes. 

This observation has a second consequence. The non-perturbative superpotential 
is not a sum of exponential terms generated
by gaugino condensates and membrane instantons, a structure which is
not in any case expected to appear in the effective supergravity. 
Instead, we find a non-perturbative term which is the exponential of a
sum of terms linear in the chiral fields, a structure characteristic of
threshold corrections induced by anomaly-cancellation 
in higher dimensions.  

In our reduction scheme, the ``microscopic" superpotential $c+\alpha M^3$ is 
moduli-independent. It is however known that T-duality induces a holomorphic dependence
on $T$ compatible with our supermultiplet structure as described in 
Lagrangian (\ref{Lcond2}). The existence of dual descriptions of moduli $S$ and $\hat S$
in terms of constrained vector multiplets $V$ and $\hat V$ or in terms of linear multiplets
$L$ and $\hat L$ implies that the ``microscopic" superpotential cannot depend on $S$ or 
$\hat S$. 

\subsection{The scalar potential}\label{secpotential}

Because of the mixing of the three moduli multiplets $S$, $T$ and $\hat S$
in the K\"ahler metric, the scalar potential present in the component expansion of the 
effective Lagrangian (\ref{Lcond2}) is not positive and analysing its 
vacuum structure is a severe problem. This mixing is due to the kinetic terms of the 
five-brane massless modes, it is unavoidable whenever five-branes, and then 
whenever $\hat S$, are present. 

We may however gain insight by deriving the scalar potential directly in 
terms of the components of the constrained vector multiplets. 
This version of the theory is indeed more 
tractable than the chiral one since the mixing of 
moduli fields is simpler. The relevant multiplets are then $T$, $V$, $\hat{V}$ for the
moduli and $U$ for the condensate. Charged matter terms are as before omitted since 
we are interested in vacua where they vanish.
In the Einstein frame (\ref{z0is}), the relation between the dilaton (\ref{dilaton}) and the 
lowest scalar component $C$ of $V$ is
\beq 
\label{dilaton2}
\varphi = \frac{1}{4\kappa^2_4 C} 
\eeq
and the scalar potential is eventually expressed in terms of the physical dilaton.
As explained in eqs. (\ref{Simplies}), eliminating $S$ with $\Omega_a$ omitted generates 
the modified Bianchi identity  $U = -8\Sigma(V)$, where $V$ is the vector field 
describing $L-2\Omega_0|_{cond.}$.
To derive the potential, we only need the scalar components of $V$ and $U$,
\beq
\begin{array}{ll}
V =(C,0,H,K,0,0,d), \qquad\qquad
& U = (u,0,f_U),  
\crbig
u=-4(H-iK), \qquad\qquad & f_U = 4d.
\end{array}
\eeq
Since a non-zero condensate $u$ also switches on the field $H-iK$ of the
dilaton multiplet $V$, the gaugino condensate clearly breaks supersymmetry 
in this sector, as expected in a theory where the dilaton couples to gauge fields. 
Eliminating $\hat S$ defines the five-brane (effective) linear multiplet 
$V\hat V = \hat L_0 - 2\hat\beta_0 V$ and, since linear multiplets do not have auxiliary 
fields, we may simply write
$$
\hat L_0 = ( \, C\hat C + 2\hat\beta_0 C \, ,0,0,0,0,0,0 \, ),
$$
in terms of the lowest scalar component $\hat C$ of $\hat V$, 
when deriving the scalar potential. The resulting scalar potential is then a function of 
the physical scalars $C$ (the dilaton, see. eq. \ref{dilaton2}), the
$S^1/Z_2$ modulus $\hat C = y/(2\pi\kappa)$, the Calabi-Yau volume 
modulus $T$ and the gaugino condensate $u$. It is also a function of the 
auxiliary fields $d$, $f_T$ and $f_0$ 
(in the compensating multiplet $S_0$) which can be easily eliminated. 

The K\"ahler potential $\widetilde K$ generates a term quadratic in $d$. 
We will write the scalar potential by restricting $\widetilde K$ to its leading term \cite{VY}
\beq
\label{Kahl}
\widetilde K (US_{0}^{-3},\overline{US}_{0}^{-3})S_{0}\ov{S}
_{0}=A(U\ov{U})^{1/3}\,,
\eeq
with an arbitrary normalisation constant $A \geq 0$. 
The scalar potential as a function of $d$, $C$, $\hat C$, $T$ and $u$ reads then:
\beq
\label{Veff}
\begin{array}{rcl}
V_{eff.} &=& {\ds-\frac{32}{9}A(u\ov{u})^{-2/3}d^{2}-Bd } \crbig&&
+ {\displaystyle \frac{u\ov{u}}{16C} \left[ {3\over2} \frac{1}{4\kappa_4^2C}
+ \tau (\hat{C}+2\hat{\beta}_0)^{2}(T+\ov{T})\right] }
\crbig
&& \displaystyle
+ \frac{2\kappa_4^2 C}{\kappa_4^4 (T+\ov{T})^{3}} \Biggl\{
- 2 \left| W + {1\over8}\kappa_4^3 \, u \left({ T+\ov T \over 2\kappa_4^2 C}\right)^{3/2}
\right|^2
\crbig
&&  \displaystyle  +\frac{(T+\ov{T})^{2}}{3}\left| 
W_{T}-\frac{3}{T+\ov{T}}W
+{1\over4}\kappa_4^3 \, u {(T+\ov T)^{3/2} \over (2\kappa_4^2 C)^{1/2}}
\left( \beta _{0}+\tau [ \hat{C}^{2}+4\hat{\beta}_{0}\hat{C
}] \right) \right| ^{2} \Biggr\} \,. 
\end{array}
\eeq
The first two terms arise respectively from the condensate 
K\"ahler potential term $[S_0\ov S_0 \widetilde K]_D = A[(U\ov U)^{1/3}]_D$ and 
from the condensate $F$--density 
$$
[ W_{cond.}]_F \equiv 
\Bigl[ {1\over4}(S+\beta_0T+\tau\hat\beta_0\hat S) U
+ {b_0 \over 96\pi^2} \{ U\ln(U/\mu^3) -U \}  \Bigr]_F.
$$
The coefficient $B$ relates the gaugino condensate field $u$ and the gauge coupling 
$g_0^2$, as defined in eq. (\ref{g2}):
\beq
B=\frac{2}{g_{0}^{2}}+\frac{b_{0}}{24\pi ^{2}}\ln \left( \frac{u\ov{u}}{\mu
^{6}}\right)
= 8 \Re \, {\partial\over\partial u} W_{cond.} \,.
\eeq
The standard field-theory value of the condensate, 
\beq
\label{condens}
|u| =\mu ^{3}\text{exp}\left(-\frac{24\pi ^{2}}{b_{0}g_{0}^{2}}\right) ,
\eeq
is obtained if $B=0$ is part of the scalar potential vacuum equations. 

In eq. (\ref{Veff}), the second line is proportional to $|H-iK|^2$ and the fourth line to
$|f_T|^2$, and we have included the possibility of a $T$--dependent perturbative superpotential, as often implied by $T$--duality, even if our reduction scheme predicts
$W_T = {\partial W\over\partial T}=0$.
The dependence of the potential on the five-brane position $\hat C = y/(2\pi\kappa)$
is best understood by defining the {\it distance $\Delta_c$ from the brane to the condensate}:
\beq
\begin{array}{ll}
\makebox{For a condensate on plane 1:} &\quad \Delta_c = y/\pi = 2\kappa \,\hat C.
\crbig
\makebox{For a condensate on plane 2:} &\quad \Delta_c = 1 - y/\pi = 1- 2\kappa \,\hat C.
\end{array}
\eeq
Using then the values of the threshold coefficients found in eqs. (\ref{hatbetaare})
and (\ref{2beta}), we find in both cases:
\beq
\begin{array}{rcl}
\beta_0 + \tau [ \hat C^2 + 4 \hat\beta_0\hat C ]  &=& \displaystyle
\beta_0^{(pl.)} + {\tau\over4\kappa^2}
\Bigl[ (1-\Delta_c)^2  - {1\over3} \Bigr],
\crbig
(\hat C+2\hat\beta_0)^2 &=& \displaystyle {1\over4\kappa^2} (1-\Delta_c)^2.
\end{array}
\eeq
These results agree with the dependence on the
five-brane location found in gauge couplings (\ref{g2fin}).

At this stage, we have two options. We may neglect the K\"ahler potential 
$\widetilde K$ and assume $A=0$. Then, the auxiliary field $d$ 
imposes the field equation 
$B=0$ and the correct value of the gaugino condensate. This procedure is equivalent to the 
derivation of the effective non-perturbative superpotential (\ref{Wnpis}). The field equation
$B=0$ allows to eliminate $u$ and to express the effective potential, which does not
include the first line in expression (\ref{Veff}), as a function 
of the moduli scalars $C$, $\hat C$ and $T$ only. 

Instead, with a non-zero K\"ahler potential $\widetilde K$ ({\it i.e.}~with $A > 0$), 
solving 
for the auxiliary $d$ turns the first two terms of the scalar potential (\ref{Veff}) into
\beq
\label{Bterm}
\frac{9}{128}\frac{B^{2}}{A}(u\ov{u})^{2/3}\,,
\eeq
and a generic (non supersymmetric) stationary point of the potential does not 
necessarily lead to $B=0$ and to the standard gaugino condensate (\ref{condens}).
But since $B$ appears quadratically in the potential, the same
stationary points with $B=0$ would exist in both cases $A=0$ and $A>0$. 
Notice that the condensate term (\ref{Bterm}) can also be written
$$
{1\over2} (\widetilde K_{u\ov u})^{-1} \Bigl( \Re \, {\partial\over\partial u} W_{cond.} \Bigr)^2,
\qquad\qquad
\widetilde K_{u\ov u} = {\partial^2\over\partial u \,\partial \ov u} A(u\ov u)^{1/3}.
$$
This is the potential term due to the auxiliary field $f_u$ of the condensate chiral superfield
$U$, with non-standard Weyl weight $w=3$. The imaginary part of 
${\partial\over\partial u} W_{cond.}$ does not contribute to the potential 
because of the constraint $U = -8\Sigma(V)$ \cite{BDQQ1}.

A complete analysis of the stationary values of the scalar potential (\ref{Veff}) 
cannot be performed analytically. In the absence of five-branes, the potential 
can be written in the form
$$
V = {1\over2} (\widetilde K_{u\ov u})^{-1} \Bigl(\Re {\partial\over\partial u} W_{cond.}\Bigr)^2
+{1\over\kappa_4^2} \Bigl[ (2\kappa_4^2 C)^2 |f_S|^2 + 3(T+\ov T)^{-2} |f_T|^2
\Bigr]  
- {3\over \kappa_4^4} e^{\cal K}|W|^2 ,
$$
in terms of the K\"ahler potential ${\cal K} = -\ln(S+\ov S) - 3\ln(T+\ov T)$
with diagonal metric.  
A relatively simple study of the stationary points of the potential with for instance 
$B=0$ can be performed as a function of the auxiliary fields
$f_S$ and $f_T$ of the chiral dilaton $S$ and the volume modulus $T$ respectively.

But the introduction of the five-brane mode leads to mixings of the chiral superfields.
For instance, according to the second superfield eq. (\ref{dualtransf}), the auxiliary field 
in the chiral dilaton multiplet $S$ reads
\beq
\label{fsis}
\begin{array}{rcl}
f_S &=& \kappa_4^{-1}(2\kappa_4^2C)^{-1/2}(T+\ov T)^{-3/2}
\biggl[ \ov W - {1\over4} \left( {T+\ov T \over 2\kappa_4^2 C} \right)^{3/2} \kappa_4^3 \ov u \biggr]
\crbig
&& + \tau \left[ \hat C^2 f_T - {1\over2} (T+\ov T) 
(\hat C^2 + 2\hat\beta_0\hat C) (2\kappa_4^2 C)^{-1} \kappa_4^2 \ov u \right].
\end{array}
\eeq
The second term is due to the five-brane and it involves the auxiliary field $f_T$,
which is proportional to the last line in the potential (\ref{Veff}).
Similarly, the auxiliary component of the five-brane multiplet $\hat S$ is
\beq
\label{fhatsis}
f_{\hat S} = 4\hat C f_T - (T+\ov T)(\hat C + 2\hat\beta_0) (2\kappa_4^2 C)^{-1} \kappa_4^2\ov u .
\eeq
Mixings of the auxiliary fields then arise whenever $\hat C\ne0$, {\it i.e.}~whenever the
five-brane does not lie on the fixed hyperplane at $y=0$. 

\section{Conclusion}\label{secconcl}

In this paper, we have studied the Calabi-Yau reduction of the low-energy limit
of M-theory on the interval $S^1/Z_2$, with five-branes aligned to preserve
four supercharges. We have used a fully consistent, four-dimensional 
supergravity and superfield setup and included the modulus field describing
five-brane fluctuations in the interval direction. 
The gauge anomaly-cancelling topological term is modified
as a consequence of the five-brane contributions to M-theory Bianchi identities.
We have derived the new four-dimensional interactions induced by these 
five-brane modifications and shown that they lead to new gauge
threshold corrections with a calculable dependence on the five-brane position 
along $S^1$. In particular, these threshold corrections fit nicely the
change in the instanton number expected when a five-brane
is moved onto one of the fixed hyperplanes. 
Of course, when gauge condensation occurs, these threshold corrections 
explicitly appear in the effective non-perturbative superpotential,
with an exponential dependence of the five-brane location.

The same five-brane-dependent contributions to the non-perturbative 
superpotential can be obtained from a different perspective.
It is expected that open membranes 
wrapping a two-cycle in the Calabi-Yau threefold and extending from a 
five-brane to a ten-dimensional hyperplane generate, in the four-dimensional 
effective field theory, instanton-like contributions. Instanton calculus
allows to explicitly compute these instanton corrections and the 
resulting contributions to the non-perturbative effective superpotential.
Strictly speaking however, instanton calculus only applies in specific limits,
which in the case under scrutiny 
restricts the understanding of the global
 structure of the superpotential and of the interplay of the various moduli fields. 
It is precisely here that our effective supergravity Lagrangian, as 
derived from M-theory, adds important new information. In particular, since membrane 
instanton corrections actually originate from threshold corrections related 
to ten-dimensional anomaly cancellation, the non-perturbative superpotential
is the exponential of a sum of terms linear in moduli chiral fields.

\hspace{1.5cm}
\section*{Acknowledgements}

We thank Adel Bilal and Claudio Scrucca for helpful discussions. 
This work has been supported by the Swiss National Science
Foundation and by the Commission of the European Communities
under contract MRTN-CT-2004-005104.

\hspace{1.5cm}

\section*{Appendix A: conventions and notations}\label{secapp}
\renewcommand{\theequation}{A.\arabic{equation}}
\setcounter{equation}{0}

Our conventions are as in refs.~\cite{BM} and \cite{BDS}. We use the {\it upstairs} 
picture $M_4 \times K_6 \times S^1/Z_2$, where the $S^1$ coordinate is 
$x^{10}=yR$, with a $(2\pi)$--periodic angular variable $y$. We
use $-\pi < y \leq \pi$ when explicit values are needed because of the natural action 
of $Z_2$ in this interval. 
Our indices convention for the $M_4 \times K_6 \times S^1/Z_2$ reduction is
$$
x^M = (x^A, yR) = (x^\mu, z^i, \ov z^{\ov k}, yR), \quad
M=0,\ldots,10, \quad A=0,\ldots,9, \quad i, \ov k=1,2,3.
$$

For bulk moduli, we use the terminology familiar from string compactifications: 
the moduli $s=\Re S$ and $t=\Re T$ with K\"ahler potential (\ref{usualK}) are 
respectively related to the dilaton (or string coupling) and to the Calabi-Yau volume. 
This convention follows from the metric
\beq
\label{StringFrame}
ds^2_{11} = e^{- 2\phi /3} g^{(10)}_{AB}\, dx^A dx^B + 
e^{4 \phi /3}dy^2\,,
\eeq
which defines the string frame and the string coupling $e^{-2\phi}$ with
$R = e^{2\phi/3}$, together with 
\beq
\label{StringFrame2}
g^{(10)}_{AB}\, dx^A dx^B = g_{\mu\nu} \, dx^\mu dx^\nu + 
V^{1/3} \delta_{i\bar{j}}\, dz^i d\ov z^{\bar{j}} 
\eeq
which defines the Calabi-Yau volume. Rescaling $g_{\mu\nu}$ to the 
four-dimensional Einstein frame leads to
\beq
\label{metr1}
ds^2_{11} = e^{4 \phi /3 }\, [V^{-1} g_{\mu\nu}\,dx^\mu dx^\nu + dy^2]
+ e^{-2\phi/3}V^{1/3} \delta_{i\bar{j}}\, dz^i dz^{\bar{j}}\,.
\eeq
Comparison with the standard eleven-dimensional metric used to diagonalize 
kinetic terms in the four-dimensional ${\cal N}=1$ supergravity Lagrangian,
\beq
\label{metr2}
ds^2_{11} = s^{-2/3}\, [t^{-1} g_{\mu\nu}\,dx^\mu dx^\nu + t^2 dy^2]
+ s^{1/3} \delta_{i\bar{j}}\, dz^i dz^{\bar{j}}\,,
\eeq
leads to the identifications 
\beq
(\Re T)^3 = V\,, \qquad \Re S = V e^{-2\phi}\,.
\eeq
Hence, $t$ is the volume modulus while $s$ is the dilaton or string coupling modulus. 

The terminology often used in the context of M-theory defines instead another 
Calabi-Yau volume $\hat V$ in units specified by the metric (\ref{metr1}), 
with then 
$$
\hat V \equiv V e^{-2\phi} = \Re S, \qquad \qquad
(\Re T)^3  = \hat V e^{2\phi}   = \hat V R^3.
$$
It seemingly exchanges the respective roles of the bulk moduli.

In order to avoid duplication of contributions due to $Z_2$ periodicity, our eleven-dimensional supergravity action and Green-Schwarz terms are multiplied by 1/2 with respect to standard
conventions in use for $M_{11}$:
\beq
{\cal L}_{C.J.S.} + {\cal L}_{G.S} =
{1\over4\kappa^2} \left[ eR - {1\over2} G_4\wedge*G_4
-{1\over6} C_3\wedge G_4 \wedge G_4\right] - {T_2\over4\pi} \, G_4\wedge X_7\,,
\eeq
with membrane tension $T_2 = 2\pi (4\pi\kappa^2)^{-1/3}$, and
\beq
dX_7 = X_ 8 = {1\over (2\pi)^3 4!} \left[ {1\over8} {\rm tr}\, R^4 - {1\over32} ({\rm tr}\, R^2)^2\right].
\eeq

To respect $Z_2$ symmetry, we assume that a five-brane with world-volume located 
at some $y_0$ has a $Z_2$--mirror at $-y_0$. The Dirac distribution transverse to its 
world-volume $M_4 \subset {\cal W}_6 \subset M_4\times K_6$ is then defined by the 
condition
\beq
\int_{M_4 \times K_6 \times S^1/Z_2} I_6 \wedge \delta^{(5)}({\cal W}_6)
= 2 \int_{{\cal W}_6} I_6
\eeq
for any 6--form $I_6$, since it takes both copies into account. 

The membrane and five-brane tensions $T_2$ and $T_5$ are related by the 
Dirac-Zwanziger quantization condition 
$$
2\kappa^2T_2T_5 = 2\pi
$$
and also by \cite {deA}
$$
(T_2)^2 = 2\pi T_5.
$$
We then express all constants in terms of $\kappa^2$, with
\beq 
(4\pi\kappa^2)^{1/3} = {2\pi\over T_2} = 2\kappa^2 T_5.
\eeq
With these conventions, the Bianchi identity is
\beq
dG_4 = {4\pi\over T_2} \left( {1\over2}\sum_I q_I \, \delta^{(5)}({\cal W}_{6,I})
- \sum_i \delta_i I_{4,i} \right) ,
\eeq
where the index $I$ labels the $Z_2$--symmetric pairs of five-branes and the charge
$q_I$ is $+1$ for a five-brane, $-1$ for an anti-five-brane.

One subtlety when integrating the Bianchi identity is that one cannot find a $Z_2$--odd function $\epsilon(y)$ such that $d\epsilon = \delta(y-y_0)dy$. As in ref.~\cite{BDS}, we then use $ y, y_0 \in ] -\pi, \pi ]$ and
\beq
\begin{array}{ll}
{\ds \epsilon_{y_0} (y) = {\rm sgn}\, (y-y_0) - {y-y_0 \over \pi } , \qquad\qquad }&
{\ds d\epsilon_{y_0} (y)  = \Big(2\delta(y-y_0) - {1\over\pi }\Big) dy, }
\crbig
{\ds\epsilon_1(y) = \epsilon_{0} (y) =  {\rm sgn}\, (y) - {y \over \pi } , \qquad\qquad} &
{\ds d\epsilon_1 (y)  = 2\delta_1 - {1\over\pi }dy,}
\crbig
{\ds \epsilon_1(y) = \epsilon_{\pi} (y) =  {\rm sgn}\, (y-\pi) - {y-\pi \over \pi } , \qquad\qquad} &
{\ds d\epsilon_2 (y)  = 2\delta_2 - {1\over\pi } dy.}
\end{array}
\eeq
The sign function is
\beq
{\rm sgn}\,(r) = {r\over|r|}\quad{\rm if}\quad r\ne 0, \qquad\qquad
{\rm sgn}\,(0) = 0,
\qquad \qquad {\rm sgn}\,(r) = -{\rm sgn}\,(-r).
\eeq
With this definition, $\epsilon_1$ and $\epsilon_2$ are odd functions
while $\epsilon_{y_0} (-y) = - \epsilon_{-y_0} (y)$. The function
\beq
\epsilon_{y_0} (y) + \epsilon_{-y_0} (y) = 
{\rm sgn}\,(y-y_0) + {\rm sgn}\,(y+y_0) - {2y\over\pi}
\eeq
($0< y_0 <\pi$) is then odd with
\beq
d[ \epsilon_{y_0} (y) + \epsilon_{-y_0} (y) ] = 2\Big(\delta(y-y_0) + \delta(y+y_0) \Big)dy 
- {2dy\over\pi}.
\eeq
This function is useful to insert five-brane sources in the Bianchi identity. 
Since the five-brane world-volumes ${\cal W}_{6,I}$ are of the form 
$M_4\times {\cal C}_2$  (${\cal C}_2$ a holomorphic cycle in $K_6$), 
${\cal W}_{6,I}$ is located at $y=y_I$ with
a  ``$Z_2$--mirror five-brane" at $-y_I$. We then use 
\beq
\begin{array}{l}
{\ds\delta^{(5)}({\cal W}_{6,I}) = \Big( \delta(y-y_I) + \delta(y+y_I)\Big) dy
\wedge \delta^{(4)}({\cal W}_{6,I}),}
\crbig
{\ds d\Big([ \epsilon_{y_I}(y) + \epsilon_{-y_I}(y) ] \,  \delta^{(4)}({\cal W}_{6,I}) \Big) =
2\,\delta^{(5)}({\cal W}_{6,I}) - {2\over\pi}\, dy \wedge \delta^{(4)}({\cal W}_{6,I}) ,}
\end{array}
\eeq
to integrate five-brane contributions to the Bianchi identity.


\end{document}